\documentclass[lettersize,journal]{IEEEtran}
\usepackage{booktabs}
\usepackage{diagbox}
\usepackage{float}
\usepackage[utf8]{inputenc}
\usepackage{balance}
\usepackage{textcomp}
\usepackage{stfloats}
\usepackage[english]{babel}
\usepackage{graphicx}
\usepackage[figurename=Fig.]{caption}
\usepackage{subcaption}
\usepackage{framed}
\usepackage[linesnumbered,ruled,vlined,commentsnumbered]{algorithm2e}
\usepackage[normalem]{ulem}
\usepackage{arydshln}
\usepackage{mathtools}
\usepackage{amsmath}
\usepackage{amsthm}
\usepackage{amssymb}
\usepackage{amsfonts}
\usepackage{dsfont}
\usepackage{enumerate}
\usepackage{comment}
\usepackage[mathlines]{lineno}
\usepackage[sorting=none]{biblatex}
\usepackage{csquotes}
\usepackage{wasysym}
\usepackage{bm}
\usepackage{hyperref}
\hypersetup{
    colorlinks=true,
    linkcolor=blue,
    filecolor=magenta,      
    urlcolor=cyan,
}
\newtheorem{theorem}{Theorem}[section]
\newtheorem{lemma}[theorem]{Lemma}
\newtheorem{proposition}[theorem]{Proposition}

\newtheorem{corollary}[theorem]{Corollary}
\newtheorem{assume}[theorem]{Assumption}
\newcommand{\rn}{\mathbb{R}^n}
\newcommand{\rN}{\mathbb{R}^N}
\newcommand{\Rmn}{\mathbb{R}^{m\times n}}
\newcommand{\RNn}{\mathbb{R}^{N\times n}}
\newcommand{\st}{\mathcal{S}}\newcommand{\sgn}{\mathrm{sign}}

\newcommand{\opnorm}{2\rightarrow2}
\begin{document}

\title{DECONET: an Unfolding Network\\ for Analysis-based Compressed Sensing\\ with Generalization Error Bounds}
\author{Vicky Kouni
\thanks{V. Kouni acknowledges financial support for the implementation of this paper by Greece and the European Union (European Social Fund-ESF) through the Operational Program ``Human Resources Development, Education and Lifelong Learning'' in the context of the Act ``Enhancing Human Resources Research Potential by undertaking a Doctoral Research'' Sub-action 2: IKY Scholarship Program for PhD candidates in the Greek Universities}
\thanks{V. Kouni and Y. Panagakis are with Dep. of Informatics \& Telecommunications, National \& Kapodistrian University of Athens, Athens, Greece. Emails: vicky-kouni@di.uoa.gr and yannisp@di.uoa.gr\\ This paper has supplementary downloadable material available at \url{http://ieeexplore.ieee.org}, provided by the authors. The material includes a pdf file, which contains proofs of some theoretical results and link to a code repository for reproducibility purposes. This material is 134KB in size}and Yannis Panagakis}

\markboth{Journal of \LaTeX\ Class Files,~Vol.~18, No.~9, September~2020}%
{V. Kouni and Y. Panagakis: DECONET: an Unfolding Network for Analysis-based Compressed Sensing with Generalization Error Bounds}

\maketitle

\begin{abstract}
    We present a new deep unfolding network for analysis-sparsity-based Compressed Sensing. The proposed network coined Decoding Network (DECONET) jointly learns a decoder that reconstructs vectors from their incomplete, noisy measurements and a redundant sparsifying analysis operator, which is shared across the layers of DECONET. Moreover, we formulate the hypothesis class of DECONET and estimate its associated Rademacher complexity. Then, we use this estimate to deliver meaningful upper bounds for the generalization error of DECONET. Finally, the validity of our theoretical results is assessed and comparisons to state-of-the-art unfolding networks are made, on both synthetic and real-world datasets. Experimental results indicate that our proposed network outperforms the baselines, consistently for all datasets, and its behaviour complies with our theoretical findings.
\end{abstract}

\section{Introduction}\label{introsec}
\emph{Compressed Sensing} (CS) \cite{cs} is a modern technique to reconstruct signals $x\in\mathbb{R}^n$ from few linear and possibly corrupted observations $y=Ax+e\in\mathbb{R}^m$, $m<n$. Iterative methods applied on CS are by now widely used \cite{daubechies,irls,boyd:5}. Nevertheless, deep neural networks (DNNs) have become popular for tackling sparse recovery problems like CS \cite{dagan,dr2}, since they significantly reduce the time complexity and increase the quality of the reconstruction. A new line of research lies on merging DNNs and optimization algorithms, leading to the so-called \emph{deep unfolding/unrolling} \cite{unrolling,unfolding}. The latter pertains to interpreting the iterations of well-known iterative algorithms as layers of a DNN, which reconstructs $x$ from $y$.
\par Deep unfolding networks have become increasingly popular in the last few years \cite{lista}, \cite{wcs}, \cite{bertocchi}, \cite{scarlett} because -- in contrast with traditional DNNs -- they are interpretable \cite{interp}, integrate prior knowledge about the signal structure \cite{deligiannis}, and have relatively few trainable parameters \cite{adm}. Especially in the case of CS, unfolding networks have proven to work particularly well. For example, \cite{ampnet}, \cite{optista}, \cite{ista-net}, \cite{tfrnn}, \cite{admm-net}, \cite{fpc}, \cite{istagen} interpret the iterations of well-studied optimization algorithms as layers of a neural network, which learns a \emph{decoder} for CS, i.e., a function that reconstructs $x$ from $y$. Additionally, some of these networks jointly learn a \emph{sparsifying transform} for $x$. This sparsifier may either be a nonlinear transform \cite{ista-net} or an orthogonal matrix \cite{istagen} -- integrating that way a dictionary learning technique. The latter has shown promising results when employed in model-based CS \cite{imagedl,iot,1bitcs}; hence, it looks appealing to combine it with unfolding networks. Furthermore, research community focuses lately on the \textit{generalization error} \cite{shalev,infogen} of deep unfolding networks \cite{listagen}, \cite{istagen}, \cite{deeprnn}, \cite{unfoldrnn}. Despite recent results of this kind, estimating the generalization error of unfolding networks for CS is still in its infancy.
\par In fact, generalization error bounds are only provided for unfolding networks that promote \textit{synthesis sparsity} in CS, by means of the dictionary learning technique. On the other hand, the \emph{analysis sparsity model} differs significantly \cite{cosparsity} from its synthesis counterpart and it can be more advantageous for CS \cite{analvssyn}. For example, the redundancy of so-called analysis operators can lead to a more flexible sparse representation of the signals of interest, compared to orthogonal sparsifying transforms (see Sections \ref{synsparse} and \ref{analvssyn} for a detailed comparison between the two sparsity models). To the best of our knowledge, only one unfolding network  \cite{admmdad} takes advantage of analysis sparsity in CS, in terms of learning a redundant sparsifying analysis operator. Nevertheless, the generalization ability of \cite{admmdad} is not mathematically explained.
\par In this paper, we are inspired by the articles \cite{admm-net}, \cite{istagen},
\cite{listagen}, \cite{deeprnn}, \cite{admmdad}. These publications propose ISTA-based \cite{daubechies}, reweighted FISTA-based \cite{fista} and ADMM-based \cite{boyd:5} unfolding networks, which jointly learn a decoder for CS and a sparsifying transform. Particularly, the learnable sparsifiers of \cite{admm-net,istagen,listagen,deeprnn} promote synthesis sparsity, while \cite{admmdad} employs its handier analysis counterpart. The deficiency of \cite{admm-net,admmdad} lies on the fact that their proposed frameworks are not accompanied by a generalization analysis, whereas \cite{istagen,listagen,deeprnn} provide generalization error bounds for the proposed networks. Similarly, we develop a new unfolding network based on an optimal analysis-$l_1$ algorithm \cite{tfocs:6} and call it \emph{Decoding Network} (DECONET). The latter jointly learns a decoder for CS and a redundant sparsifying analysis operator; thus, we address the CS problem under the analysis sparsity model. Our novelty lies on estimating the generalization error of the proposed analysis-based unfolding network. To that end, we upper bound the generalization error of DECONET in terms of the Rademacher complexity \cite{resrad} of the associated hypothesis class. In the end, we numerically test the validity of our theory and compare our proposed network to the state-of-the-art (SotA) unfolding networks of \cite{istagen} and \cite{admmdad}, on real-world and synthetic data. In all datasets, our proposed neural architecture outperforms the baselines in terms of generalization error, which scales in accordance with our theoretical results.
\par Our key contributions are listed below.
\begin{enumerate}
    \item After differentiating synthesis from analysis sparsity in CS and presenting example unfolding networks in Section \ref{relwork}, we develop in Section \ref{acfsec} a new unfolding network dubbed DECONET. The latter jointly learns a decoder that solves the analysis-based CS problem and a redundant sparsifying analysis operator $W\in\mathbb{R}^{N\times n}$, $n<N$, that is shared across the layers of DECONET.
    \item We introduce in Section \ref{mainsec} the hypothesis class -- parameterized by $W$ -- of all the decoders DECONET can realize and restrict $W$ to be bounded in this class, so that we impose a realistic structural constraint on the operator.
    \item Later in Section \ref{mainsec}, we estimate the generalization error of DECONET using a chaining technique. Our results showcase that the redundancy $N$ of $W$ and the number of layers $L$ affect the generalization ability of DECONET; roughly speaking, the generalization error scales like $\sqrt{NL}$ (see Theorem \ref{gengentheorem} and Corollary \ref{asymptoticL}). To the best of our knowledge, we are the first to study the generalization ability of an unfolding network for analysis-based CS.
    \item We confirm the validity of our theoretical guarantees in Section \ref{expsec}, by testing DECONET on a synthetic dataset and two real-world image datasets, i.e., MNIST \cite{mnist} and CIFAR10 \cite{cifar}. We also compare DECONET to two SotA unfolding networks: a recent variant of ISTA-net \cite{istagen} and ADMM-DAD net \cite{admmdad}. Our experiments demonstrate that a) the generalization error of DECONET scales correctly with our theoretical findings b) DECONET outperforms both baselines, consistently for all datasets.
\end{enumerate}

\begin{table*}[t!]
    \centering
    \scalebox{1.0}{\begin{tabular}{||c|c|c|c||}
    \hline
    DUN & Iterative Scheme & Sparsity Model & Gen. Error Bounds\\
    \hline\hline
    ISTA-net \cite{istagen} & Iterative Soft Thresholding Algorithm \cite{daubechies} & Synthesis & Yes\\
    \hline
    SISTA-RNN \cite{tfrnn} & Sequential Iterative Soft Thresholding Algorithm \cite{sista} & Synthesis & No\\
    \hline
    Reweighted-RNN \cite{deeprnn} & Reweighted $l_1-l_1$ algorithm \cite{deeprnn} & Synthesis & Yes\\
    \hline
    AMP-Net \cite{ampnet} & Approximate Message Passing \cite{ampalgo} & Synthesis & No\\
    \hline
    ADMM-net \cite{admm-net} & Alternating Direction Method of Multipliers \cite{boyd:5} & Synthesis & No\\
    \hline
    ADMM-DAD \cite{admmdad} & Alternating Direction Method of Multipliers \cite{boyd:5} & Analysis & No\\
    \hline
    DECONET (proposed) & Analysis-$l_1$ \cite{tfocs:6} & Analysis & Yes\\
    \hline
    \end{tabular}}
    \caption{Comparisons among some example unfolding networks for CS. Categorizations are based on a) the associated optimization algorithm b) the employed sparsity model c) the study of generalization error.}
    \label{unfoldingnets}
\end{table*}

\par\textbf{Notation.} We denote the set of real, positive numbers by $\mathbb{R}_{+}$. For a sequence $a_n$ that is upper bounded by $M>0$, we write $\{a_n\}\leq M$. For a matrix $A\in\mathbb{R}^{n\times n}$, we write $\|A\|_{\opnorm}$ for its operator/spectral norm and $\|A\|_F$ for its Frobenius norm. Moreover, we write $\|A\|_{\opnorm}\approx1$ if $\|A\|_{\opnorm}\geq1$, but there exists $C>0$ such that $C\|A\|_{\opnorm}\leq1$. For a family of vectors $(w_i)_{i=1}^N$ in $\rn$, its associated analysis operator is given by $W f:=\{\langle f,w_i\rangle\}_{i=1}^N$, where $f\in\rn$. For square matrices $A_1,A_2\in\mathbb{R}^{N\times N}$, we denote by $[A_1;A_2]\in\mathbb{R}^{2N\times N}$ their concatenation with respect to the first dimension, while we denote by $[A_1\,|\,A_2]\in\mathbb{R}^{N\times 2N}$ their concatenation with respect to the second dimension. Similarly, for non-square matrices $A_1\in\mathbb{R}^{m_1\times n}$,  $A_2\in\mathbb{R}^{m_2\times n}$, $A_3\in\mathbb{R}^{m\times n_1}$,  $A_4\in\mathbb{R}^{m\times n_2}$, we denote by $[A_1;A_2]\in\mathbb{R}^{(m_1+m_2)\times n}$ the concatenation of $A_1$ and $A_2$ with respect to the first dimension, while we denote by $[A_3\,|\,A_4]\in\mathbb{R}^{m\times (n_1+n_2)}$ the concatenation of $A_3$ and $A_4$ with respect to the second dimension. We write $O_{N\times N}$ for a real-valued $N\times N$ matrix filled with zeros and $I_{N\times N}$ for the $N\times N$ identity matrix. We denote by $\mathrm{diag}(\alpha)$ a square diagonal matrix having $\alpha\in\mathbb{R}$ in its main diagonal and zero elsewhere. For $x\in\mathbb{R},\,\tau>0$, the soft thresholding operator $\st_\tau:\mathbb{R}\mapsto\mathbb{R}$ is defined as
    \begin{equation}\label{sto}
    \st_\tau(x)=\st(x,\tau)=
    \begin{cases}
    \sgn(x)(|x|-\tau),& |x|\geq\tau\\
    0,&\text{otherwise,}
    \end{cases}
    \end{equation} or in closed form $\st(x,\tau)=\sgn(x)\max(0,|x|-\tau)$. For $x\in\rn$, the soft thresholding operator acts component-wise, i.e. $(\st_\tau(x))_i=\st_\tau(x_i)$, and is 1-Lipschitz with respect to $x$. For $y\in\rn,\,\tau>0$, the mapping
    \begin{equation}\label{proxmap}
        P_G(\tau;y)=\mathrm{argmin}_{x\in\rn}\left\{\tau G(x)+\frac{1}{2}\|x-y\|_2^2\right\},
    \end{equation}
    is the \textit{proximal mapping associated to the convex function G}. For $G(\cdot)=\|\cdot\|_1$, \eqref{proxmap} coincides with \eqref{sto}. For $x\in\mathbb{R},\,\tau>0$, the truncation operator $\mathcal{T}_\tau:\mathbb{R}\mapsto\mathbb{R}$ is defined as
    \begin{equation}
    \label{trunc}
        \mathcal{T}_\tau(x)=\mathcal{T}(x,\tau)=
        \begin{cases}
        \tau\sgn(x),& |x|\geq\tau\\
        x,&\text{otherwise}
        \end{cases},
    \end{equation}
    or in closed form $\mathcal{T}(x,\tau)=\sgn(x)\min\{|x|,\tau\}$. For $x\in\rn$, the truncation operator acts component-wise and is 1-Lipschitz with respect to $x$. For two functions $f,g:\mathbb{R}^n\mapsto\mathbb{R}^n$, we write their composition as $f\circ g:\mathbb{R}^n\mapsto\mathbb{R}^n$ and if there exists some constant $C>0$ such that $f(x)\leq Cg(x)$, then we write $f(x)\lesssim g(x)$. For the ball of radius $t>0$ in $\rn$ with respect to some norm $\|\cdot\|$, we write $B_{\|\cdot\|}^n(t)$. The covering number $\mathcal{N}(T,\|\cdot\|,t)$ of a space $T$, equipped with a norm $\|\cdot\|$, at level $t>0$, is defined as the smallest number of balls $B_{\|\cdot\|}^n(t)$ required to cover $T$. The set of all matrices $W\in\RNn$ with  operator norm bounded by some $0<\Lambda<\infty$, is defined as $\mathcal{B}_\Lambda=\{W\in\RNn:\,\|W\|_{\opnorm}\leq\Lambda\}$.
    
\section{Related work:\\ From model-based to data-driven CS}
\label{relwork}
The main idea of CS is to reconstruct a vector $x\in\rn$ from measurements $y=Ax+e\in\mathbb{R}^m$, $m<n$, where $A$ is the so-called \emph{measurement matrix} \cite{rf} and $e\in\mathbb{R}^m$, with $\|e\|\leq\varepsilon$, corresponds to noise. In order to ensure exact/approximate reconstruction of $x$, CS relies on two principles. First, $A$ must meet some conditions, for example the restricted isometry property or the null space property \cite{rf}. In particular, random Gaussian matrices $A\in\Rmn$ have proven to be nice candidates for CS, since they satisfy such conditions \cite{rf}. Second, we assume $x$ is (approximately) sparse. Sparse data models are split in synthesis and analysis sparsity \cite{figu}.

\subsection{Synthesis Sparsity in CS and ISTA-based Unfolding}\label{synsparse}
Under the synthesis sparsity model \cite{rf,fistashear,adcock,eeg}, signals are considered to be sparse when \emph{synthesized} by a few column vectors taken from a large matrix called \emph{dictionary}, which is typically assumed to be orthogonal, i.e. $D\in\mathbb{R}^{n\times n}$, with $DD^T=I_{n\times n}$ (e.g. $D$ may be the discrete cosine transform), so that $x=Dz$. Employing synthesis sparsity in CS, we aim to recover $x$ from $y$. A common way to do so is by solving the \emph{$l_1$-minimization} problem
\begin{equation}\label{l1syn}
    \min_{z\in\mathbb{R}^n}\|z\|_1\quad\text{s. t.}\quad\|y-ADz\|_2\leq\varepsilon.
\end{equation}
Towards that end, numerous iterative algorithms \cite{daubechies,irls,ipm} have emerged. Typically, they consist of an iterative scheme that incorporates a proximal mapping and after a number of iterations and under certain conditions, they converge to a minimizer $\hat{x}$ of \eqref{l1syn}. For example, ISTA uses the proximal mapping \eqref{proxmap} to yield the following iterative scheme
\begin{align}
    z_{k+1}=&\st_{\tau\lambda}(z_k+\tau(AD)^T(y-ADz))\notag\\
    \label{istaalgo}
    =&\st_{\tau\lambda}((I-D^TA^TAD)z+\tau(AD)^Ty),
\end{align}
for $k=0,1,\dots$, $z_0=0$, with $\tau,\lambda>0$ being parameters of the algorithm. If $\tau\|AD\|_{\opnorm}^2\leq1$ \cite{daubechies}, $z_k$ converges to a minimizer $\hat{z}$ of \eqref{l1syn}, so that the reconstructed $\hat{x}$ is simply given by $\hat{x}=D\hat{z}$. As stated in \cite{istagen}, under the assumption that $D$ is learned from a training set, the iterative scheme of \eqref{istaalgo} can be interpreted as a layer of a neural network (whose trainable parameters are the entries of $D$) with weight matrix $I-D^TA^TAD$, bias $\tau(AD)^Ty$ and activation function $\st_{\tau\lambda}$. Then, the composition of a given number of layers and the consequent application of $D$ constitutes the decoder implemented by the ISTA-based network, which outputs $\hat{x}\approx x$.
\subsection{Analysis Sparsity in CS and ADMM-based Unfolding}\label{analvssyn}
Despite its success, synthesis sparsity has a ``twin'', i.e., the \emph{analysis sparsity model} \cite{kr,genzel,cosparse}, in which one assumes that there exists a \emph{redundant analysis operator} $W\in\RNn$, $n<N$, so that  $Wx$ is sparse. For example, $W$ may be the analysis operator associated to a \emph{frame} \cite{grochenig,feichtinger} or a finite difference operator \cite{tv}. The associated optimization problem for CS is the \emph{analysis $l_1$-minimization} problem
\begin{equation}\label{l1anal}
        \min_{x\in\rn}\|W x\|_1\quad\text{s. t.}\quad\|Ax-y\|_2\leq\varepsilon.
\end{equation}
From now on, whenever we speak about the \emph{redundancy} of an analysis operator, we mean the number of its rows $N$.\\
Analysis sparsity has become popular, due to some benefits it has compared to its synthesis counterpart. For example, it is computationally more appealing to solve the optimization algorithm of analysis-based CS, since the actual optimization takes place in the ambient space \cite{star} and the algorithm involved may need less measurements for perfect reconstruction, if one uses a redundant transform instead of an orthogonal one \cite{genzel}. Nevertheless, choosing the appropriate iterative algorithm for solving \eqref{l1anal} may be a tricky task. The reason is that most thresholding algorithms cannot handle analysis sparsity, since the proximal mapping associated to $\|W(\cdot)\|_1$ does not have a closed-form type. To tackle this issue and solve \eqref{l1anal}, one may employ the so-called ADMM \cite{boyd:5} algorithm, which uses the following iterative scheme
\begin{align}
\label{xup}
    &x^{k+1}=(A^TA+\rho W^TW)^{-1}(A^Ty+\rho W^T(z^k-u^k))\\
    \label{zup}
     &z^{k+1}=\st_{\lambda/\rho}(W x^{k+1}-u^k)\\
        \label{uup}
        &u^{k+1}=u^k+W x^{k+1}-z^{k+1},
\end{align}
with $k\in\mathbb{N}$, dual variables $z,u\in\mathbb{R}^N$, initial points $(x^0,z^0,u^0)=(0,0,0)$, penalty parameter $\rho>0$ and regularization parameter $\lambda>0$. As shown in \cite{boyd:5}, the iterates \eqref{xup} -- \eqref{uup} converge to a solution $p^\star$ of
\begin{equation}
    \min_{x\in\rn}\frac{1}{2}\|Ax-y\|_2^2+\lambda\|z\|_1\quad\text{s. t.}\quad Wx-z=0.
\end{equation}
In \cite{admmdad}, the updates \eqref{xup} - \eqref{uup} are  formulated
as a neural network (whose trainable parameters are the entries of $W$) with $L$ layers, defined as 
\begin{align}
    g_1(y) & =I_1b(y)+I_2\st_{\lambda/\rho}(b(y)),\\
    g_k(v) & =\Tilde{\Theta}v+I_1b+I_2\st_{\lambda/\rho}(\Theta v+b), \quad k = 2, \hdots, L,
\end{align}
with $v\in\mathbb{R}^{2N\times 1}$ being an \emph{intermediate variable}, weight matrices $\Tilde{\Theta}$, $\Theta$ (depending on $A$, $W$, $\rho$), bias $b$ (depending on $A$, $W$, $\rho$, $y$) and activation function $\st_{\lambda/\rho}$. The resulting learned decoder that reconstructs $x$ from $y$ emerges by applying an affine map on the composition of a given number of layers.
\section{DECONET: a New Unfolding Network\\ for Analysis-based CS}\label{acfsec}
The generic ADMM-based unfolding network of \cite{admmdad} seems promising, but it involves the costly computation of $(A^TA+\rho W^TW)^{-1}$. On the other hand, ADMM-based unfolding networks may deal with the bottleneck of the inverses' computation, by leveraging some structure of the problem \cite{snapshot} or imposing specific constraints \cite{admm-csnet}. Nevertheless, such unfolding networks are synthesis-based and thus they cannot treat analysis sparsity. Therefore, in order to address both aspects, we opt for solving \eqref{l1anal} with the optimal first-order \emph{analysis}-$l_1$ algorithm described in \cite{tfocs:6}, which uses transposes -- instead of inverses -- of the involved matrices. We will briefly describe the steps leading to the derivation of the aforementioned algorithm, as these are stated in \cite{tfocs:6}. First, an equivalent to \eqref{l1anal} \emph{smoothed} formulation is given as follows:
\begin{equation}\label{l1conic}
    \min_{x\in\rn}\|Wx\|_1+\frac{\mu}{2}\|x-x_0\|_2^2\quad\text{s. t.}\quad\|y-Ax\|_2\leq\varepsilon,
\end{equation}
where $\mu\in\mathbb{R}_{+}$ is the so-called \emph{smoothing parameter} and $x_0\in\rn$ is an initial guess on $x$. Second, the dual of \eqref{l1conic} is determined to be
\begin{equation}\label{dual}
\begin{split}
     \text{max}_{z^2\in\mathbb{R}^m}\quad&\langle y,z^2\rangle-\varepsilon\|z^2\|_2\\
\text{s. t.}\quad &A^Tz^2-W^Tz^1=0,\ \|z^1\|_{\infty}\leq1,
\end{split}
\end{equation}
where $z^1\in\rN$, $z^2\in\mathbb{R}^m$ are dual variables. The afore-described formulations -- combined with a collection of arguments and computations -- lead to Algorithm \ref{l1analysis}, which constitutes a variant of an optimal first-order method. For reasons of convenience, we call this algorithm \emph{analysis conic form} (ACF) from now on. ACF also involves step sizes $\{t_k^1\},\,\{t_k^2\}>0$ and a step size multiplier $0<\{\theta_k\}$. A standard setup for ACF employs update rules such that $0<\{t_k^1\}_{k\geq0}$, $\{\,t_k^2\}_{k\geq0}\leq1$, $0<\{\theta_k\}_{k\geq0}\leq1$.

\begin{algorithm}[h!]

\SetAlgoLined
\SetKwData{Left}{left}\SetKwData{This}{this}\SetKwData{Up}{up}\SetKwFunction{Union}{Union}\SetKwFunction{FindCompress}{FindCompress}\SetKwInOut{Input}{Input}\SetKwInOut{Output}{Output}
\Input{$x_0\in\rn,\,z_0^1\in\rN,\,z_0^2\in\mathbb{R}^m$, $\mu\in\mathbb{R}_+$, step sizes $\{t_k^1\},\{\,t_k^2\}$}
\Output{solution $\hat{x}_{\mu}$ of \eqref{l1conic}}
$\theta_0\leftarrow1,\,u_0^1=z_0^1,\,u_0^2=z_0^2$\;
\For{$\mathrm{iterations} \ k=0,1,\dots$}{
$x_k\leftarrow x_0+\mu^{-1}((1-\theta_k)W^Tu_k^1+\theta_kW^Tz_k^1-(1-\theta_k)A^Tu_k^2-\theta_kA^Tz_k^2)$\;
$z_{k+1}^1\leftarrow\mathcal{T}((1-\theta_k)u_k^1+\theta_kz_k^1-\theta_k^{-1}t_k^1Wx_k,\theta_k^{-1}t_k^1)$\;
$z_{k+1}^2\leftarrow\st((1-\theta_k)u_k^2+\theta_kz_k^2-\theta_k^{-1}t_k^2(y-Ax_k),\theta_k^{-1}t_k^2\varepsilon)$\;
$u_{k+1}^1\leftarrow(1-\theta_k)u_k^1+\theta_kz_{k+1}^1$\;
$u_{k+1}^2\leftarrow(1-\theta_k)u_k^2+\theta_kz_{k+1}^2$\;
$\theta_{k+1}\leftarrow2/(1+(1+4/(\theta_k)^2)^{1/2})$\;
}
\caption{ACF}
\label{l1analysis}
\end{algorithm}
The dual function $g_\mu$ corresponding to \eqref{dual} has a Lipschitz continuous gradient, hence ACF converges \cite{tfocs:6} to a solution $\hat{x}_\mu$ of \eqref{l1conic}, for which we have $\hat{x}_{\mu}\overset{\mu\rightarrow0}{\longrightarrow}\hat{x}$, where $\hat{x}$ is an optimal solution of \eqref{l1anal}. The authors of \cite{tfocs:6} clarify that when they speak about the optimal solution $\hat{x}$, they refer to this uniquely determined value. Additionally, they argue that there are situations where $\hat{x}$ and $\hat{x}_\mu$ coincide. Henceforward, we stick to their formulation and speak about the solution $\hat{x}$.
\par We consider a standard scenario for the ACF, where $z_0^1=u_0^1=0$, $z_0^2=u_0^2=0$, $t^1_0=t^2_0=\theta_0=1$, $0<\{t_k^1\},\{\,t_k^2\},\{\theta_k\}\leq1$, $\mu>1$, $x_0=A^Ty$ and $A\in\Rmn$ is an appropriately normalized random matrix (which constitutes a typical choice for CS), with $\|A\|_{\opnorm}\approx1$. We substitute first $x-$update into $z^1-$ and $z^2-$updates and second $z^1-$ and $z^2-$ into $u^1-$ and $u^2-$updates, respectively, concatenate $z_k^1,z_{k}^2,u_{k}^1,u_{k}^2$ in one vector $v_{k}$, i.e. $v_k^T=(z_{k}^1,z_{k}^2,u_{k}^1,u_{k}^2)^T\in\mathbb{R}^{(1\times p)}$, $p=2N+2m$, for $k\geq0$, with $v_0=0$, and do the calculations, so that
\begin{align}
\label{finalv1}
v_{k+1}&=D_{k}v_{k}+\Theta_{k}\begin{pmatrix*}[l]
\mathcal{T}(G_{k}^1v_{k}-b_{k}^1,\theta_{k}^{-1}t_{k}^1)\\
\st(G_{k}^2v_{k}-b_{k}^2,\theta_{k}^{-1}t_{k}^2\varepsilon)\\
\mathcal{T}(G_{k}^1v_{k}-b_{k}^1,\theta_{k}^{-1}t_{k}^1)\\
\st(G_{k}^2v_{k}-b_{k}^2,\theta_{k}^{-1}t_{k}^2\varepsilon)\\
\end{pmatrix*},
\end{align}
where
\begin{align}
\label{bias1}
b_k^1&=\theta_k^{-1}t_k^1Wx_0\in\mathbb{R}^N,\\
\label{bias2}
b_k^2&=\theta_k^{-1}t_k^2(y-Ax_0)\in\mathbb{R}^m,\\
\label{thetak}
\Theta_k&=\mathrm{diag}(\underbrace{\theta_0,\dots,\theta_0}_{(N+m)\ \text{times}},\underbrace{\theta_k,\dots,\theta_k}_{(N+m)\ \text{times}})\in\mathbb{R}^{p\times p},\\
\label{dk}
D_k&=(I_{p\times p}-\Theta_k)\in\mathbb{R}^{p\times p},
\end{align}
\begin{align}
G_k^1=&(\theta_k(I-\theta_k^{-1}t_k^1\mu^{-1}WW^T)\mid t_k^1\mu^{-1}WA^T\notag\\\label{g1}
&\mid(1-\theta_k)(I-\theta_k^{-1}t_k^1\mu^{-1}WW^T)\\
&\mid(1-\theta_k)\theta_k^{-1}t_k^1\mu^{-1}WA^T)\in\mathbb{R}^{N\times p},\notag\\
G_k^2=&(t_k^2\mu^{-1}AW^T\mid\theta_k(I-\theta_k^{-1}t_k^2\mu^{-1}AA^T)\notag\\
&\mid(1-\theta_k)\theta_k^{-1}t_k^2\mu^{-1}AW^T\mid(1-\theta_k)\notag\\
\label{g2}
&\cdot(I-\theta_k^{-1}t_k^2\mu^{-1}AA^T))\in\mathbb{R}^{m\times p}.
\end{align}

We observe that \eqref{finalv1} can be interpreted as a layer of a neural network, with weights $G^1,\,G^2$, biases $b^1,\,b^2$ and activation functions $\mathcal{T}(\cdot,\cdot),\,\st(\cdot,\cdot)$. Nevertheless, this interpretation of ACF as a DNN does not account for any trainable parameters. We cope with this issue by considering $W$ to be unknown and learned from a training sequence $\mathbf{S}=\{(x_i,y_i)\}_{i=1}^s$ with i.i.d. samples drawn from an unknown distribution\footnote{Formally speaking, this is a distribution over the $x_i$ and then $y_i=Ax_i+e$, with fixed $A,e$} $\mathcal{D}^s$. Hence, the trainable parameters are the entries of $W$. Additionally, we make the realistic assumption that $W$ is bounded with respect to the operator norm, i.e., $W\in\mathcal{B}_\Lambda$. Based on \eqref{finalv1}, we formulate ACF as a neural network with $L$ layers/iterations, defined as
\begin{align}
    \label{layer1}
    f_1(y)&=\sigma_1(y)\\
    \label{layerk}
    f_k(v)&=D_{k-1}v+\Theta_{k-1}\sigma_{k-1}(v),\quad k=2,\dots,L,
\end{align}
where
\begin{align}
        \sigma_1(y)^T=(&\mathcal{T}(-t_0^1Wx_0,t_0^1),\mathcal{S}(t_0^2(y-Ax_0),t_0^2\varepsilon),\notag\\
        \label{ysigma}
        &\mathcal{T}(-t_0^1Wx_0,t_0^1),\mathcal{S}(t_0^2(y-Ax_0),t_0^2\varepsilon))^T,\\
        \sigma_k(v)^T=(&\mathcal{T}(G_k^1v-b_k^1),\mathcal{S}(G_k^2v-b_k^2),\notag\\ \label{vsigma}
        &\mathcal{T}(G^1_kv-b^1_k),\mathcal{S}(G_k^2v-b_k^2))^T,
    \end{align}
for $k=2,\dots,L$. We denote the composition of $L$ such layers (all having the same $W$) as
\begin{equation}
\label{composed}
    f_{W}^L(y)=f_L\circ f_{L-1}\circ\dots\circ f_1(y).
\end{equation}
The latter constitutes the realization of a neural network with $L$ layers, that reconstructs the intermediate variable $v$ from $y$. Thus, we call \eqref{composed} \emph{intermediate decoder}. Motivated by the $x$-update of ACF, we apply an affine map $\phi:\mathbb{R}^{p\times 1}\mapsto\mathbb{R}^{n\times1}$ after the last layer $L$, yielding the desired solution $\hat{x}$:
\begin{equation}\label{affinemap}
    \hat{x}:=\phi(v)=\Phi v+x_0,
\end{equation}
where
\begin{equation}
    \begin{split}
    \label{phi}
        \Phi=(&\mu^{-1}\theta_LW^T\vert-\mu^{-1}\theta_LA^T\vert\ \mu^{-1}(1-\theta_L)W^T\\
        &\vert-\mu^{-1}(1-\theta_L)A^T)\in\mathbb{R}^{n\times p}.
    \end{split}
\end{equation}
Moreover, in order to clip the output $\phi(f_W^L(y))$ in case its norm falls out of a reasonable range, we apply an extra function $\psi : \mathbb{R}^n \to \mathbb{R}^n$ after $\phi$ and define it as
\begin{equation}\label{psi}
    \psi(x) = \left\{\begin{array}{cc}
    x,&\ \|x\|_2 \leq B_{\mathrm{out}}\\
    B_{\mathrm{out}}\frac{x}{\|x\|_2},&\ \text{otherwise}
    \end{array}\right.,
\end{equation} for some constant $B_{\mathrm{out}}>0$.
For fixed $L$, the desired learnable decoder is written as
\begin{equation}\label{decoder}
    \text{dec}_W^L(y)=\psi(\phi(f^L_W(y))).
\end{equation}
We call Decoding Network (DECONET) the network that implements such a decoder, which is parameterized by $W$.

\begin{figure}[t!]
    \centering
    \includegraphics[width=\linewidth]{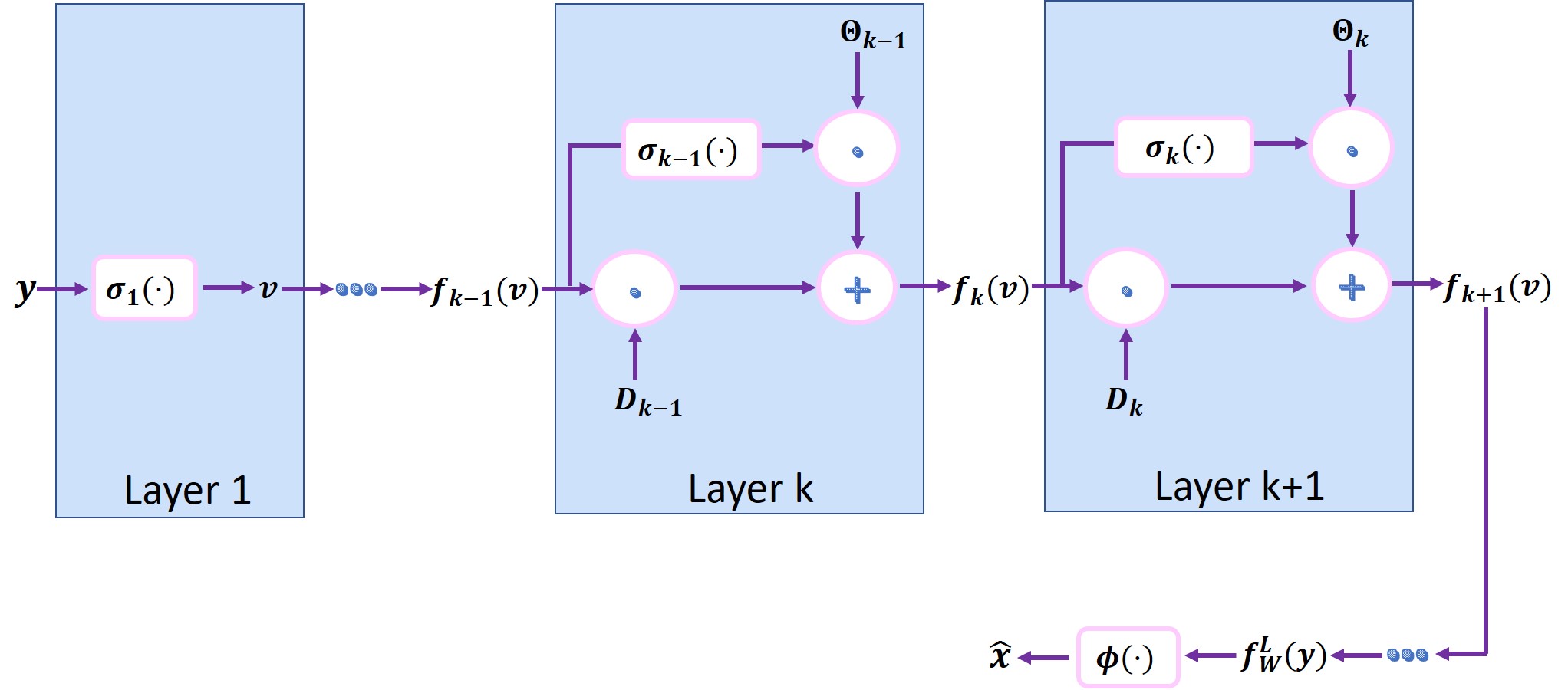}
    \caption{Schematic illustration of \eqref{layer1}, \eqref{layerk}, \eqref{composed} and \eqref{affinemap}. Note that $y$ is passed as input into every subsequent layer $k=2,\dots,L$, but for notational simplification, we only write the intermediate variable $v$. The operators `` $\pmb{\cdot}$ '' and `` $\pmb{+}$ '' denote matrix-vector multiplication and vector-vector addition, respectively, while the activation functions $\sigma_1(\cdot)$ and $\sigma_{k}(\cdot)$ ($k=2,\dots,L$) are defined as in \eqref{ysigma} and \eqref{vsigma}, respectively.}
    \label{dec-arc}
\end{figure}
\section{Generalization Analysis of DECONET}\label{mainsec}
In this Section, we deliver meaningful -- in terms of $L$ and $N$ -- upper bounds on the generalization error of DECONET. We do so in a series of steps presented in the next subsections.
\subsection{Hypothesis Class of DECONET and Associated Rademacher Complexity}\label{hyposec}
We introduce the hypothesis class
\begin{equation}\label{hypothesis}
    \mathcal{H}^L=\{ h:\mathbb{R}^m\mapsto\rn:\, h(y)=\psi(\phi(f^L_{W}(y))),\,W\in\mathcal{B}_\Lambda\},
\end{equation}
parameterized by $W$ and consisting of all the functions/decoders DECONET can implement. Given \eqref{hypothesis} and the training set $\mathbf{S}$, DECONET yields a function $h_{\mathbf{S}}\in\mathcal{H}^L$ that aims at reconstructing $x$ from $y$. For a loss function $\ell:\mathcal{H}^L\times\rn\times\mathbb{R}^m\mapsto\mathbb{R}_+$, the empirical loss of a hypothesis $h\in\mathcal{H}^L$ is the reconstruction error on the training set, i.e.
\begin{equation}\label{empiricalrisk}
    \mathcal{\hat{L}}_{train}(h)=\frac{1}{s}\sum_{i=1}^s\ell(h,x_i,y_i).
\end{equation}
In this paper, we choose as loss function $\ell$ the squared $l_2$-norm; hence, \eqref{empiricalrisk} takes the form of the training \emph{mean-squared error} (MSE):
\begin{equation}\label{trainmse}
    \mathcal{\hat{L}}_{train}(h)=\frac{1}{s}\sum_{j=1}^s\|h(y_j)-x_j\|_2^2.
\end{equation}
The true loss is
\begin{equation}
    \mathcal{L}(h)=\mathbb{E}_{(x,y)\sim\mathcal{D}}(\|h(y)-x\|_2^2).
\end{equation}
The generalization error is given as the difference\footnote{Some of the existing literature denotes the true loss as the generalization error, but the definition we give in \eqref{generror} is more convenient for our purposes} between the empirical and true loss
\begin{equation}\label{generror}
    \text{GE}(h)=|\mathcal{\hat{L}}_{train}(h)-\mathcal{L}(h)|.
\end{equation}

A typical way to estimate \eqref{generror} consists in upper bounding it in terms of the \emph{Rademacher complexity} \cite[Definition 2]{rademacher}. The \emph{empirical Rademacher complexity} is defined as \begin{equation}
    \label{erc}
    \mathcal{R}_{\mathbf{S}}(\ell\circ\mathcal{H}^L)=\mathbb{E}\sup_{h\in\mathcal{H}^L}\frac{1}{s}\sum_{i=1}^s\epsilon_i\|h(y_i)-x_i\|_2^2,
\end{equation}
where $\epsilon$ is a Rademacher vector, that is, a vector with i.i.d. entries taking the values $\pm1$ with equal probability. Then, the Rademacher complexity is defined as
\begin{equation}\label{rademacher}
    \mathcal{R}_s(\ell\circ\mathcal{H}^L)=\mathbb{E}_{{\mathbf{S}}\sim\mathcal{D}^s}(\mathcal{R}_{\mathbf{S}}(\ell\circ\mathcal{H}^L)).
\end{equation}
In this paper, we solely work with \eqref{erc}. We rely on the following Theorem that estimates \eqref{generror} in terms of \eqref{erc}.
\begin{theorem}[{\cite[Theorem 26.5]{shalev}}]\label{radem}
Let $\mathcal{H}$ be a family of functions, $\mathbf{S}$ the training set drawn from $\mathcal{D}^s$, and $\ell$ a real-valued bounded loss function satisfying $|\ell(h,z)|\leq c$, for all $h\in\mathcal{H}, z\in Z$. Then, for $\delta\in(0,1)$, with probability at least $1-\delta$, we have for all $h\in\mathcal{H}$
\begin{equation}
    \mathcal{L}(h)\leq\mathcal{\hat{L}}(h)+2\mathcal{R}_{\mathbf{S}}(\ell\circ\mathcal{H})+4c\sqrt{\frac{2\log(4\delta)}{s}}.
\end{equation}
\end{theorem}
In order to use the previous Theorem, the loss function must be bounded. Towards this end, we make two typical -- for the machine learning literature -- assumptions regarding the training set $\mathbf{S}=\{(x_i,y_i)\}_{i=1}^s$. Let us suppose that  with overwhelming probability we have
\begin{equation}
    \label{ybound}
    \|y_i\|_2\leq\mathrm{B_{in}},
\end{equation}
for some constant $\mathrm{B_{in}}>0$, $i=1,\dots,s$. Moreover, we assume that for any $h\in\mathcal{H}^L$, with overwhelming probability over $y_i$ chosen from $\mathcal{D}$, the following holds
\begin{equation}\label{hbound}
    \|h(y_i)\|_2\leq B_{\text{out}},
\end{equation}
by definition of $\psi$, for some constant $B_{\text{out}}>0$, for all $i=1,\dots,s$. Hence, $\|\cdot\|_2^2$ is bounded as $\|h(y_i)-x_i\|_2^2\leq (B_{\text{in}}+B_{\text{out}})^2$, $i=1,\dots,s$. Following the previous assumptions, it is easy to check that $\|\cdot\|^2_2$ is a Lipschitz continuous function, with Lipschitz constant $\mathrm{Lip}_{\|\cdot\|^2_2}=2\mathrm{B_{in}}+2\mathrm{B_{out}}$; hence, we can employ the so-called (vector-valued) contraction principle, which allows us to study $\mathcal{R}_\mathbf{S}(\mathcal{H})$ alone: 
\begin{lemma}[{\cite[Corollary 4]{contraction}}]
\label{contraction}
Let $\mathcal{H}$ be a set of function $h:\mathcal{X}\mapsto\mathbb{R}^n$, $f:\mathbb{R}^n\mapsto\mathbb{R}^n$ a $K$-Lipschitz function and $\mathbf{S}=\{x_i\}_{i=1}^s$. Then
\begin{equation}
    \mathbb{E}\sup_{h\in\mathcal{H}}\sum_{i=1}^s\epsilon_if\circ h(x_i)\leq\sqrt{2}K\mathbb{E}\sup_{h\in\mathcal{H}}\sum_{i=1}^s\sum_{k=1}^n\epsilon_{ik}h_k(x_i),
\end{equation}
where $(\epsilon_i),(\epsilon_{ik})$ are both Rademacher sequences.
\end{lemma}
Applying Lemma \ref{contraction} in $\mathcal{R}_{\mathbf{S}}(\ell\circ\mathcal{H})$ yields:
\begin{align}
    \mathcal{R}_\mathbf{S}(l\circ\mathcal{H}^L)\leq&\sqrt{2}\mathrm{Lip}_{\|\cdot\|^2_2}\mathcal{R}_\mathbf{S}(\mathcal{H}^L)\notag\\
    \label{contradem}
    =&\sqrt{2}\mathrm{Lip}_{\|\cdot\|^2_2}\mathbb{E}\sup_{h\in\mathcal{H}^L}\sum_{i=1}^s\sum_{k=1}^n\epsilon_{ik}h_k(x_i)\notag\\
    =&\mathrm{B}_{\text{in}}^\text{out}\mathbb{E}\sup_{h\in\mathcal{H}^L}\sum_{i=1}^s\sum_{k=1}^n\epsilon_{ik}h_k(x_i),
\end{align}
where $\mathrm{B}_{\text{in}}^\text{out}=\sqrt{2}(2\mathrm{B_{in}}+2\mathrm{B_{out}})$. We return to estimating \eqref{contradem} in Section \ref{coversec}, after presenting the adequate mathematical tools in Sections \ref{boundsec} and \ref{lipsec}.
\subsection{Boundedness of DECONET's Outputs}\label{boundsec}
We take into account the number of training samples and pass to matrix notation. Due to \eqref{ybound} and the Cauchy-Schwartz inequality, we get $\|Y\|_F\leq\sqrt{s}\mathrm{B}_{\text{in}}$. Similarly, the application of the Cauchy-Schwartz inequality in \eqref{hbound} yields
\begin{align}
\label{boundedout}
\|h(Y)\|_F&=\|\psi(\phi(f_{W}^L(Y)))\|_F\leq\sqrt{s}\mathrm{B}_{\text{out}}.
\end{align}

\begin{assume}
Since there exist redundant analysis operators $W$ for which $\Lambda$ may be relatively small \cite[Corollary 6.2.3.]{grochenig}, \cite[Section V]{framebound}, \cite[Proposition 5.1]{faulhuber}, \cite[Lemma 1]{rdwt}, we may reasonably assume that $c_{1,k}\Lambda\leq1$ and $c_{1,k}\Lambda^2\leq1$, for all $k\geq0$. This simplifying assumption will hold for the remainder of this paper.
\end{assume}

The next Lemma presents bounds on a quantity we will encounter more often in the sequel.

\begin{lemma}[Proof in the supplementary material]\label{g12bound}
Let $k\geq0$. For any $W\in\mathcal{B}_\Lambda$, step sizes $0<\{t_k^1\},\{t_k^2\}\leq1$ with $t_0^1=t_0^2=1$, step size multiplier $0<\{\theta_k\}\leq1$ with $\theta_0=1$, and smoothing parameter $\mu>1$, the following holds for the matrices $G^1_k,\,G_k^2$ defined in \eqref{g1}, \eqref{g2}, respectively:
\begin{equation}
    \begin{split}\label{g1+2}
        2\|G_k^1\|_{\opnorm}+2\|G_k^2\|_{\opnorm}+1\leq\Gamma_k,
    \end{split}
\end{equation}
where
\begin{equation}
    \begin{split}
        \Gamma_k=2\big[c_{1,k}\Lambda^2&+c_{2,k}\|A\|_{\opnorm}^2\\
        &+2\|A\|_{\opnorm}\Lambda(c_{1,k}+c_{2,k})\big]+1,
    \end{split}
\end{equation}
with $\{c_{1,k}\}=\{\theta^{-1}_{k}\mu^{-1}t^1_{k}\}\leq1$, $\{c_{2,k}\}=\{\theta^{-1}_{k}\mu^{-1}t^2_{k}\}\leq1$, for all $k\geq0$. Moreover, if $c_{1,k}\Lambda\leq1$, $c_{1,k}\Lambda^2\leq1,\,c_{2,k}\|A\|_{\opnorm}^2\leq1$, then $\Gamma_k\leq\gamma$ for all $k\geq0$, with $\gamma=4(\Lambda+\|A\|_{\opnorm}+1)+1$.
\end{lemma}
Apart from \eqref{boundedout}, we can upper-bound the output $f^k_W(Y)$ with respect to the Frobenius norm, after any number of layers $k$ and especially for $k<L$, so that $\phi$ and $\psi$ are not applied after the final layer $L$.
\begin{lemma}\label{fbound}
Let $k\in\mathbb{N}$. For any $W\in\mathcal{B}_\Lambda$, step sizes $0<\{t_k^1\}$, $\{t_k^2\}\leq1$ with $t_0^1=t_0^2=1$, $t_{-1}^1=t_{-1}^2=0$, step size multiplier $0<\{\theta_k\}\leq1$ with $\theta_0=\theta_{-1}=1$, and smoothing parameter $\mu>1$, the following holds for the output of the functions $f_W^k$ defined in \eqref{layer1} - \eqref{layerk}:
\begin{equation}\label{finductive}
    \|f^k_W(Y)\|_F\leq2\mu\|Y\|_F\Bigg(\sum_{i=0}^{k-1}\left(Q_{i-1}\prod_{j=i}^{k-1}\Gamma_j\right)+Q_{k-1}\Bigg),
\end{equation}\\
where $Q_k=\|A\|_{\opnorm}(c_{1,k}\Lambda+c_{2,k}\|A\|_{\opnorm})+ c_{2,k}$, $k\geq0$, with $Q_{-1}=0$ and $\{\Gamma_k\}_{k\geq0}$, $\{c_{1,k}\}_{k\geq0}$, $\{c_{2,k}\}_{k\geq0}\leq1$ are defined as in Lemma \ref{g12bound}. Moreover, if $c_{1,k}\Lambda\leq1$, $c_{1,k}\Lambda^2\leq1,\,c_{2,k}\|A\|_{\opnorm}^2\leq1$, then we have the simplified upper bound
\begin{equation}
    \|f^k_W(Y)\|_F\leq2\mu\|Y\|_F(\|A\|_{\opnorm}+1)(\zeta_k+1),
\end{equation}
where $\zeta_k=\frac{\gamma^k-1}{\gamma-1}$, with $\gamma$ defined as in Lemma \ref{g12bound}.
\end{lemma}
\begin{proof}
By definition of \eqref{sto} and \eqref{trunc}, $\mathcal{S}(\cdot,\cdot)$ and $\mathcal{T}(\cdot,\cdot)$ are 1-Lipschitz functions with respect to their first parameter, respectively. For the matrices $\Theta_k$ and $D_k$ in \eqref{thetak} and \eqref{dk} respectively, we have $\|D_k\|_{2\rightarrow2}\leq1$ and $\|\Theta_k\|_{2\rightarrow2}=1$, for any $k\geq1$, since $0<\{\theta_k\}\leq1$ and $\theta_0=1$. We use the previous statements, along with \eqref{layer1} and \eqref{layerk}, to prove \eqref{finductive} via induction. For $k=1$, we have
\begin{align*}
    \|f_W^1(Y)\|_F\leq&2t_0^1\Lambda\|X_0\|_F+2t_0^2(\|Y\|_F+\|A\|_{\opnorm}\|X_0\|_F)\\
    =&2\mu c_{1,0}\Lambda\|A\|_{\opnorm}\|Y\|_F\\
    &+2\mu c_{2,0}(\|Y\|_F+\|A\|^2_{\opnorm}\|Y\|_F)\\
    =&2\mu\|Y\|_F\big(\|A\|_{\opnorm}(c_{1,0}\Lambda+c_{2,0}\|A\|_{\opnorm})+c_{2,0}\big).
\end{align*}
Suppose \eqref{finductive} holds for $k$. Then, for $k+1$:
\begin{align*}
    \|f_W^{k+1}(Y)\|_F<&\|f_W^k(Y)\|_F+2\|G_k^1f_W^k(Y)-B_k^1\|_F\\
    &+2\|G_k^2f_W^k(Y)-B_k^2\|_F\\
    \leq&\|f_W^k(Y)\|_F(2\|G_k^1\|_{\opnorm}+2\|G_k^2\|_{\opnorm}+1)\\
    &+2(\|B_k^1\|_F+\|B_k^2\|_F)\\
    \leq&\Gamma_k\|f_W^k(Y)\|_F+2\mu\|X_0\|_F(c_{1,k}\Lambda\\
    &+c_{2,k}\|A\|_{\opnorm})+2\mu c_{2,k}\|Y\|_F\\
    \leq&\Gamma_k2\mu\|Y\|_F\sum_{i=0}^{k-1}\Bigg(\bigg(Q_{i-1}\bigg)\prod_{j=i}^{k-1}\Gamma_j\Bigg)\\
    &+\Gamma_k2\mu\|Y\|_FQ_{k-1}+2\mu\|Y\|_FQ_k\\
    =&2\mu\|Y\|_F\Bigg[\sum_{i=0}^{k}\Bigg(Q_{i-1}\prod_{j=i}^{k}\Gamma_j\Bigg)+Q_k\Bigg],
\end{align*}
where in the forth inequality we set $Q_k=\|A\|_{\opnorm}(c_{1,k}\Lambda+c_{2,k}\|A\|_{\opnorm})+ c_{2,k}$ for all $k\geq0$ and applied Lemma \ref{g12bound}.
Therefore, we proved that \eqref{finductive} holds for any $k\in\mathbb{N}$. Under the additional assumptions $c_{1,k}\Lambda\leq1$, $c_{1,k}\Lambda^2\leq1,\,c_{2,k}\|A\|_{\opnorm}^2\leq1$, for any $k\geq0$, we may apply Lemma \ref{g12bound} on \eqref{finductive}, yielding
\begin{align*}
    \|f_W^k(Y)\|_F&\leq2\mu\|Y\|_F(\|A\|_{\opnorm}+1)\Bigg(\sum_{i=0}^{k-1}\prod_{j=i}^{k-1}\gamma+1\Bigg)\\
    &=2\mu\|Y\|_F(\|A\|_{\opnorm}+1)\Bigg(\sum_{i=1}^{k}\gamma^i+1\Bigg)\\
    &=2\mu\|Y\|_F(\|A\|_{\opnorm}+1)(\zeta_k+1),
\end{align*}
with $\zeta_k=\frac{\gamma^k-1}{\gamma-1}$ and $\gamma$ defined as in Lemma \ref{g12bound}.
\end{proof}

\subsection{Lipschitzness Results}\label{lipsec}
In the next Theorem, we prove that the intermediate decoder \eqref{composed} is Lipschitz continuous with respect to $W$ and explicitly calculate the Lipschitz constants, which depend on $L$.

\begin{theorem}[Proof in the supplementary material]\label{perturbation}
Let $f^L_W$ defined as in \eqref{composed}, $L\geq2$, dictionary $W\in\mathcal{B}_\Lambda$, step sizes $0<\{t_k^1\}_{k\geq0},\{t_k^2\}_{k\geq0}\leq1$ with $t_0^1=t_0^2=1$, $t_{-1}^1=t_{-1}^2=0$, step size multiplier $0<\{\theta_k\}_{k\geq0}\leq1$ with $\theta_0=\theta_{-1}=1$, and smoothing parameter $\mu>1$. Then, for any $W_1,W_2\in\mathcal{B}_\Lambda$, we have
\begin{equation}\label{fpert}
    \|f^L_{W_1}(Y)-f^L_{W_2}(Y)\|_F\leq K_L\|W_1-W_2\|_{2\rightarrow2},
\end{equation} where
\begin{align}
    K_L=&2\mu\|Y\|_F\Bigg[\mu^{-1}\|A\|_{\opnorm}+\sum_{k=2}^L\Bigg(\left(\max_{0\leq l\leq L-1}\Gamma_l\right)^{L-k}\notag\\
    &\cdot\sum_{i=0}^{k-2}2\Bigg(Q_{i-1}\prod_{j=i}^{k-2}\Gamma_j\Bigg)+2Q_{k-1}(2\Lambda c_{1,k-1}+\|A\|_{\opnorm}\notag\\
    \label{kl}
    &\cdot(c_{1,k-1}+c_{2,k-1}))+c_{1,k-1}\|A\|_{\opnorm}\Bigg)\Bigg],
\end{align}
with $\{Q_k\}_{k\geq0}$ ($Q_{-1}=0$) defined as in Lemma \ref{fbound}, $\{\Gamma_k\}_{k\geq0}$, $\{c_{1,k}\}_{k\geq0}$, $\{c_{2,k}\}_{k\geq0}$ defined as in Lemma \ref{g12bound} and $c_{1,-1}=c_{2,-1}=0$. Moreover, if $c_{1,k}\Lambda\leq1$,  $c_{1,k}\Lambda^2\leq1,\,c_{2,k}\|A\|_{\opnorm}^2\leq1$, for all $k\geq0$, then we have the simplified upper bound
\begin{equation}
\begin{split}
    \label{klsimple}
    K_L\leq&2\mu\|Y\|_F\bigg[\|A\|_{\opnorm}(L-1+\mu^{-1})\\
    &+2(\|A\|_{\opnorm}+1)(\|A\|_{\opnorm}+3)\kappa_L\bigg],
\end{split}
\end{equation}
where
\begin{align}
\label{kappal}
    \kappa_L=\gamma^L\left(\frac{L-1}{\gamma(\gamma-1)}+\frac{\gamma(\gamma-2)}{(\gamma-1)^2}\right)-\frac{\gamma^2(\gamma-2)}{(\gamma-1)^2},
\end{align}
with $\gamma$ as in Lemma \ref{g12bound}.
\end{theorem}
We also prove below the Lipschitzness of the main decoder defined in \eqref{decoder}.

\begin{corollary}
Let $h\in\mathcal{H}^L$ defined as in \eqref{hypothesis}, $L\geq2$, and dictionary $W\in\mathcal{B}_\Lambda$. Then, for any $W_1,W_2\in\mathcal{B}_\Lambda$, we have:
\begin{multline}
    \|\psi(\phi(f_{W_2}^L(Y)))-\psi(\phi(f_{W_1}^L(Y)))\|_F\\
    \leq\mu^{-1}(\Lambda+\|A\|_{\opnorm})K_L\|W_2-W_1\|_F,
\end{multline}
with $K_L$ as in Theorem \ref{perturbation}.
\end{corollary}

\begin{proof}
By definition, $\psi$ is a 1-Lipschitz function. Moreover, as an affine map, $\phi$ is Lipschitz continuous with Lipschitz constant $\text{Lip}_{\phi}=\|\Phi\|_{\opnorm}$, with $\Phi$ defined as in \eqref{phi}. We evaluate $\|\Phi\|_{\opnorm}$:
\begin{align*}
    \|\Phi\|_{\opnorm} \leq&\mu^{-1}\theta_L\|W\|_{\opnorm}+\mu^{-1}\theta_L\|A\|_{\opnorm}\\
    &+\mu^{-1}(1-\theta_L)\|W\|_{\opnorm}+\mu^{-1}(1-\theta_L)\|A\|_{\opnorm}\\
    \leq&\mu^{-1}(\Lambda+\|A\|_{\opnorm}).
\end{align*}
Combining the previous estimate with Theorem \ref{perturbation}, we get
\begin{align*}
    \|\psi(\phi(f_{W_2}^L&(Y)))-\psi(\phi(f_{W_1}^L(Y)))\|_F\\
    &\leq\|\phi(f_{W_2}^L(Y))-\phi(f_{W_1}^L(Y))\|_F\\
    &\leq\|\Phi\|_{\opnorm}\|f_{W_2}^L(Y)-f_{W_1}^L(Y)\|_F\\
    &\leq\mu^{-1}(\Lambda+\|A\|_{\opnorm})K_L\|W_2-W_1\|_F.\qedhere
\end{align*}
\end{proof}

\subsection{Covering Numbers and Dudley's Inequality}\label{coversec}
For a fixed number of layers $L\in\mathbb{N}$, we define the set $\mathcal{M}\subset\mathbb{R}^{n\times s}$ corresponding to the hypothesis class $\mathcal{H}^L$ to be
\begin{align}
    \mathcal{M}:&=\{(h(y_1)|h(y_2)|\dots|h(y_s))\in\mathbb{R}^{n\times s}:\ h\in \mathcal{H}^{L}\}\notag\\
    &=\{\psi(\phi((f^L_W(Y)))\in\mathbb{R}^{n\times s}:\ W\in\mathcal{B}_\Lambda\}.
\end{align}
The column elements of each matrix in $\mathcal{M}$ are the reconstructions given by a decoder $h\in\mathcal{H}^L$ when applied to the measurements $y_i$. Since $\mathcal{M}$ is parameterized by $W$ like $\mathcal{H}^L$ is, we may rewrite \eqref{contradem} as
\begin{align}\label{rlh}
    \mathcal{R}_\mathbf{S}(l\circ\mathcal{H}^L)&\leq\mathrm{B}_{\text{in}}^\text{out}\mathbb{E}\sup_{M\in\mathcal{M}}\frac{1}{s}\sum_{i=1}^{s}\sum_{k=1}^n\epsilon_{ik}M_{ik}.
\end{align}
Thus, we are left with estimating the Rademacher process in the right hand side of \eqref{rlh}. The latter has subgaussian increments, hence we use Dudley's inequality \cite[Theorem 8.23]{rf}, \cite[Theorem 5.23]{dudley} to upper bound it in terms of the covering numbers of $\mathcal{M}$. Towards that end, we calculate the radius of $\mathcal{M}$, that is,
\begin{align}
    \Delta(\mathcal{M})&=\sup_{h\in\mathcal{H}^L}\sqrt{\mathbb{E}\left(\sum_{i=1}^{s}\sum_{k=1}^n\epsilon_{ik}h_k(y_i)\right)^2}\notag\\
    &\leq\sup_{h\in\mathcal{H}^L}\sqrt{\mathbb{E}\sum_{i=1}^{s}\sum_{k=1}^n\epsilon_{ik}(h_k(y_i))^2}\notag\\\label{radius}
    &\leq\sup_{h\in\mathcal{H}^L}\sqrt{\sum_{i=1}^{s}\|h(y_i)\|_2^2}\overset{\eqref{boundedout}}{\leq}\sqrt{s}B_{\mathrm{out}}.
\end{align}
With \eqref{radius} in hand, applying Dudley's inequality to \eqref{rlh} yields
\begin{equation}\label{dudleys}
\begin{split}
    \mathcal{R}_\mathbf{S}(l\circ\mathcal{H}^L)\leq&\frac{16(\mathrm{B_{in}}+\mathrm{B_{out}})}{s}\\
    &\cdot\int_0^{\frac{\sqrt{s}B_{\mathrm{out}}}{2}}\sqrt{\log\mathcal{N}(\mathcal{M},\|\cdot\|_F,\varepsilon)}d\varepsilon.
\end{split}
\end{equation}

\begin{lemma}\label{cover}
For $0<\Lambda<\infty$, the covering number of $B_{\|\cdot\|_{\opnorm}}^{N\times n}(\Lambda)=\{X\in\mathbb{R}^{N\times n}:\,\|X\|_{\opnorm}\leq \Lambda\}$ satisfies the following for any $\varepsilon>0$:
\begin{equation}
    \mathcal{N}(B_{\|\cdot\|_{\opnorm}}^{N\times n}(\Lambda),\|\cdot\|_{\opnorm},\varepsilon)\leq\left(1+\frac{2\Lambda}{\varepsilon}\right)^{Nn}.
\end{equation}
\end{lemma}
\begin{proof}
For $|\cdot|$ denoting the volume in $\mathbb{R}^{N\times n}$, the following is an adaptation of a well-known result \cite[Proposition 4.2.12]{vershynin}, connecting covering numbers and volume in $\mathbb{R}^{N\times n}$:
\begin{align*}
    \mathcal{N}(B_{\|\cdot\|_{\opnorm}}^{N\times n}(\Lambda),\|\cdot\|_{\opnorm},\varepsilon)&\leq\frac{|B_{\|\cdot\|_{\opnorm}}^{N\times n}(\Lambda)+(\frac{\varepsilon}{2})B_{\|\cdot\|_{\opnorm}}^{N\times n}(1)|}{|(\frac{\varepsilon}{2})B_{\|\cdot\|_{\opnorm}}^{N\times n}(1)|}\\
    &=\frac{|(\Lambda+\frac{\varepsilon}{2})B_{\|\cdot\|_{\opnorm}}^{N\times n}(1)|}{|(\frac{\varepsilon}{2})B_{\|\cdot\|_{\opnorm}}^{N\times n}(1)|}.
\end{align*}
Hence,
\begin{align*}
    \mathcal{N}(B_{\|\cdot\|_{\opnorm}}^{N\times n}(\Lambda),\|\cdot\|_{\opnorm},\varepsilon)&\leq\left(1+\frac{2\Lambda}{\varepsilon}\right)^{Nn}.\qedhere
\end{align*}
\end{proof}
We employ the previous Lemma, in order to estimate \eqref{dudleys}.
\begin{proposition}\label{coveringm}
The following estimate holds for the covering numbers of $\mathcal{M}$:
\begin{equation}\label{coverestim}
    \mathcal{N}(\mathcal{M},\|\cdot\|_F,\varepsilon)\leq\left(1+\frac{2\Lambda(\Lambda+\|A\|_{\opnorm})K_L}{\mu\varepsilon}\right)^{Nn}.
\end{equation}
\end{proposition}
\begin{proof}
Due to Lemma \ref{cover}, we can upper bound the covering numbers of $\mathcal{B}_\Lambda$ as follows:
\begin{equation}
    \mathcal{N}(\mathcal{B}_\Lambda,\|\cdot\|_{\opnorm},\varepsilon)\leq\left(1+\frac{2\Lambda }{\varepsilon}\right)^{Nn}.
\end{equation}
Therefore, for the covering numbers of $\mathcal{M}$ we have
\begin{align*}
    \mathcal{N}(\mathcal{M},\|\cdot\|_F,\varepsilon)&\leq\mathcal{N}(\rho K_L\mathcal{B}_\Lambda,\|\cdot\|_{\opnorm},\varepsilon)\\
    &=\mathcal{N}(\mathcal{B}_\Lambda,\|\cdot\|_{\opnorm},\varepsilon/\rho K_L)\\
    &\leq\left(1+\frac{2\Lambda \rho K_L}{\varepsilon}\right)^{Nn},
\end{align*}
where $\rho=\mu^{-1}(\Lambda+\|A\|_{\opnorm})$.
\end{proof}

\subsection{Generalization Error Bounds}\label{gensec}
We are now in position to deliver generalization error bounds for DECONET.
\begin{theorem}\label{gentheorem}
Let $\mathcal{H}^L$ be the hypothesis class defined in \eqref{hypothesis}. With probability at least $1-\delta$, for all $h\in\mathcal{H}^L$, the generalization error is bounded as
\begin{align}
    \mathcal{L}(h)\leq&\mathcal{\hat{L}}(h)+8(B_{\mathrm{in}}+B_{\mathrm{out}})B_{\mathrm{out}}\sqrt{\frac{Nn}{s}}\notag\\
    &\cdot\sqrt{\log\left(e\left(1+\frac{4\mu^{-1}\Lambda(\Lambda+\|A\|_{\opnorm})K_L}{\sqrt{s}B_{\mathrm{out}}}\right)\right)}\notag\\ 
    &+4(B_{\mathrm{in}}+B_{\mathrm{out}})^2\sqrt{\frac{2\log(4/\delta)}{s}},
\end{align}
with $K_L$ defined in \eqref{kl}.
\end{theorem}
\begin{proof}
We apply Proposition \ref{coveringm} to \eqref{dudleys}, yielding:
\begin{align*}
    \mathcal{R}_\mathbf{S}(l&\circ\mathcal{H}^L)\\
    \leq&\frac{16(\mathrm{B_{in}}+\mathrm{B_{out}})}{s}\int_0^{\frac{\sqrt{s}B_{\mathrm{out}}}{2}}\sqrt{\log\mathcal{N}(\mathcal{M},\|\cdot\|_F,\varepsilon)}d\varepsilon\\
    \leq&\frac{16(\mathrm{B_{in}}+\mathrm{B_{out}})}{s}\\    &\cdot\int_0^{\frac{\sqrt{s}B_{\mathrm{out}}}{2}}\sqrt{Nn\log\left(1+\frac{2\Lambda(\Lambda+\|A\|_{\opnorm})K_L}{\mu\varepsilon}\right)}d\varepsilon\\
    \leq&8(\mathrm{B_{in}}+\mathrm{B_{out}})B_{\mathrm{out}}\sqrt{\frac{Nn}{s}}\\
    &\cdot\sqrt{\log\left(e\left(1+\frac{4\Lambda(\Lambda+\|A\|_{\opnorm})K_L}{\mu\sqrt{s}B_{\mathrm{out}}}\right)\right)},
\end{align*}
where in the last step we used the inequality\footnote{The interested reader may refer to \cite[Lemma C.9]{rf} for a detailed proof of this inequality}
\begin{equation*}
    \int_0^a\sqrt{\log\left(1+\frac{b}{t}\right)}dt\leq a\sqrt{\log(e(1+b/a))},\qquad a,b>0.
\end{equation*}
The proof follows by employing Theorem \ref{radem} with the upper bound $c=(B_{\mathrm{in}}+B_{\mathrm{out}})^2$ for the loss function $\|\cdot\|_2^2$. 
\end{proof}

If we further assume that it holds $\{c_{1,k}\Lambda\}_{k\geq0}\leq1$, $\{c_{1,k}\Lambda^2\}_{k\geq0}\leq1$, $\{c_{2,k}\|A\|_{\opnorm}^2\}_{k\geq0}\leq1$, for $\{c_{1,k}\}_{k\geq0}=\{t^1_k/\mu\theta_k\}_{k\geq0}\leq1$, $\{c_{2,k}\}_{k\geq0}=\{t^2_k/\mu\theta_k\}_{k\geq0}\leq1$, we obtain:

\begin{corollary}
\label{gencor}
Let $\mathcal{H}^L$ be the hypothesis class defined in \eqref{hypothesis} and assume that $c_{1,k}\Lambda\leq1$, $c_{1,k}\Lambda^2\leq1$, $c_{2,k}\|A\|^2_{\opnorm}\leq1$, for all $k\geq0$, with $\{c_{1,k}\}$, $\{c_{1,k}\}\leq1$ defined as in Lemma \ref{g12bound}. With probability at least $1-\delta$, for all $h\in\mathcal{H}^L$, the generalization error is bounded as
\begin{align}
    \mathcal{L}(h)\leq&\mathcal{\hat{L}}(h)+8(B_{\mathrm{in}}+B_{\mathrm{out}})\Bigg(B_{\mathrm{out}}\sqrt{\frac{Nn}{s}}\notag\\
    &\cdot\sqrt{\log\left(e\left(1+\frac{\|Y\|_F(p+qL+r\kappa_L)}{\sqrt{s}B_{\mathrm{out}}}\right)\right)}\notag\\
    &+\sqrt{\frac{2\log(4/\delta)}{s}}\Bigg),
\end{align}
with $\kappa_L$ as in Theorem \ref{perturbation} and $p,\,q,\,r>0$ constants depending on $\|A\|_{\opnorm}, \Lambda, \mu$. 
\end{corollary}
\begin{proof}
The estimate easily follows from Theorems \ref{perturbation} and \ref{gentheorem}, if we set $p:=\Lambda(\Lambda+\|A\|_{\opnorm})\|A\|_{\opnorm}$, $q:=p(\mu^{-1}-1)$ and $r:=2p(\|A\|_{\opnorm}+1)(\|A\|_{\opnorm}+3)$.
\end{proof}
All the previous results are summarized in
\begin{theorem}\label{gengentheorem}
Let $\mathcal{H}^L$ be the hypothesis class defined in \eqref{hypothesis}. Assume there exist pair-samples $\{(x_i,y_i)\}_{i=1}^s$, with $y_i=Ax_i+e$, $\|e\|_2\leq\varepsilon$, for some $\varepsilon>0$, that are drawn i.i.d. according to an unknown distribution $\mathcal{D}$, and that it holds $\|y_i\|_2\leq\mathrm{B_{in}}$ almost surely with $\mathrm{B_{in}}=\mathrm{B_{out}}$ in \eqref{psi}. Let us further assume that for step sizes $0<\{t^1_k\}_{k\geq0}$, $\{t^2_k\}_{k\geq0}\leq1$, step size multiplier $0<\{\theta_k\}_{k\geq0}\leq1$ and smoothing parameter $\mu>1$, we have $\mu^{-1}\theta_k^{-1}t^1_k\Lambda\leq1$, $\mu^{-1}\theta_k^{-1}t^1_k\Lambda^2\leq1$, $\mu^{-1}\theta_k^{-1}t^2_k\|A\|_{\opnorm}\leq1$, for all $k\geq0$. Then with probability at least $1-\delta$, for all $h\in\mathcal{H}^L$, the generalization error is bounded as
\begin{align}
    \mathcal{L}(h)\leq&\mathcal{\hat{L}}(h)+16B^2_{\mathrm{out}}\sqrt{\frac{Nn}{s}}\notag\\
    \label{genbound}
    &\cdot\sqrt{\log\left(e\left(1+\frac{\|Y\|_F(p+qL+r\kappa_L)}{\sqrt{s}B_\mathrm{out}}\right)\right)}\\
    &+16B_{\mathrm{out}}\sqrt{\frac{2\log(4/\delta)}{s}}\notag,
\end{align}
with $\kappa_L$ as in Theorem \ref{perturbation} and constants $p,\,q,\,r>0$ as in Corollary \ref{gencor}. 
\end{theorem}

We also state below a key remark, regarding the generalization error bound as a function of $L,N$ and $s$.

\begin{corollary}[Informal]
\label{asymptoticL}
According to \eqref{kappal}, we have that $L$ enters at most exponentially in the definition of $\kappa_L$. If we consider the dependence of the generalization error bound \eqref{genbound} only on $L,N,s$ and treat all other terms as constants, we roughly have
\begin{align}\label{genbigoh}
    |\mathcal{L}(h)-\mathcal{\hat{L}}(h)|\lesssim\sqrt{\frac{NL}{s}}.
\end{align}
\end{corollary}

\noindent\textbf{Comparison with related work}: Similarly to Theorem \ref{gengentheorem}, the result obtained in \cite[Theorem 2]{istagen} demonstrates that the generalization error of the proposed synthesis-sparsity-based ISTA-net roughly scales like $\sqrt{(n\log L)(n+m)/s}$. The latter estimate is slightly better in terms of $L$ than our theoretical results. On the other hand, as depicted in Theorem \ref{gengentheorem}, our bound does not depend on $m$, which makes it tighter in terms of the number of measurements.

\begin{figure*}[h!]
    \centering
    \begin{subfigure}{0.9\linewidth}
    \centering
    \includegraphics[width=\linewidth]{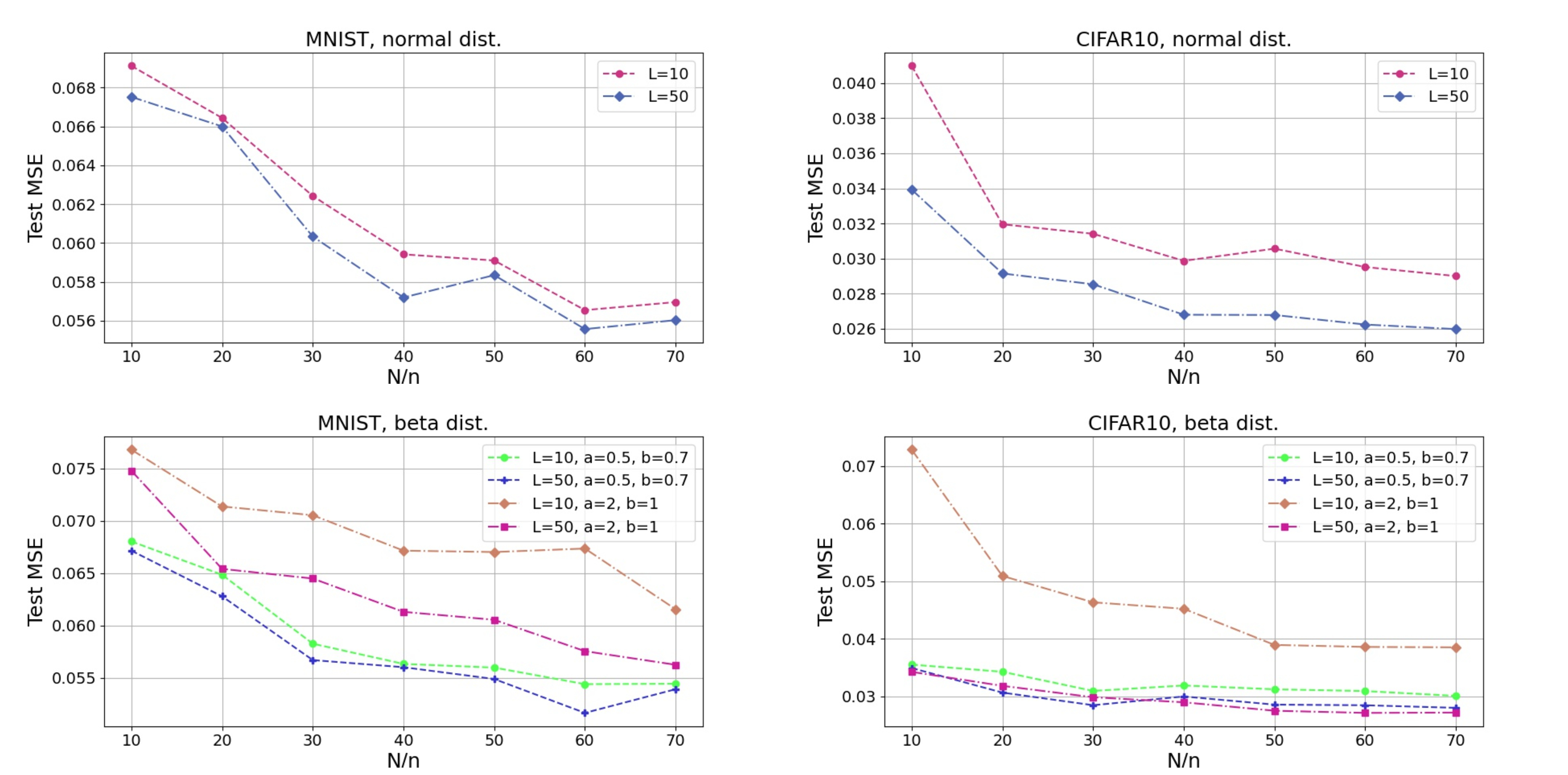}    \captionsetup{justification=centering}
    \caption{Average test MSEs with normal (top) and Beta (bottom) distributions}
    \label{testcollage}
    \end{subfigure}
    \begin{subfigure}{0.9\linewidth}
    \centering
    \includegraphics[width=\linewidth]{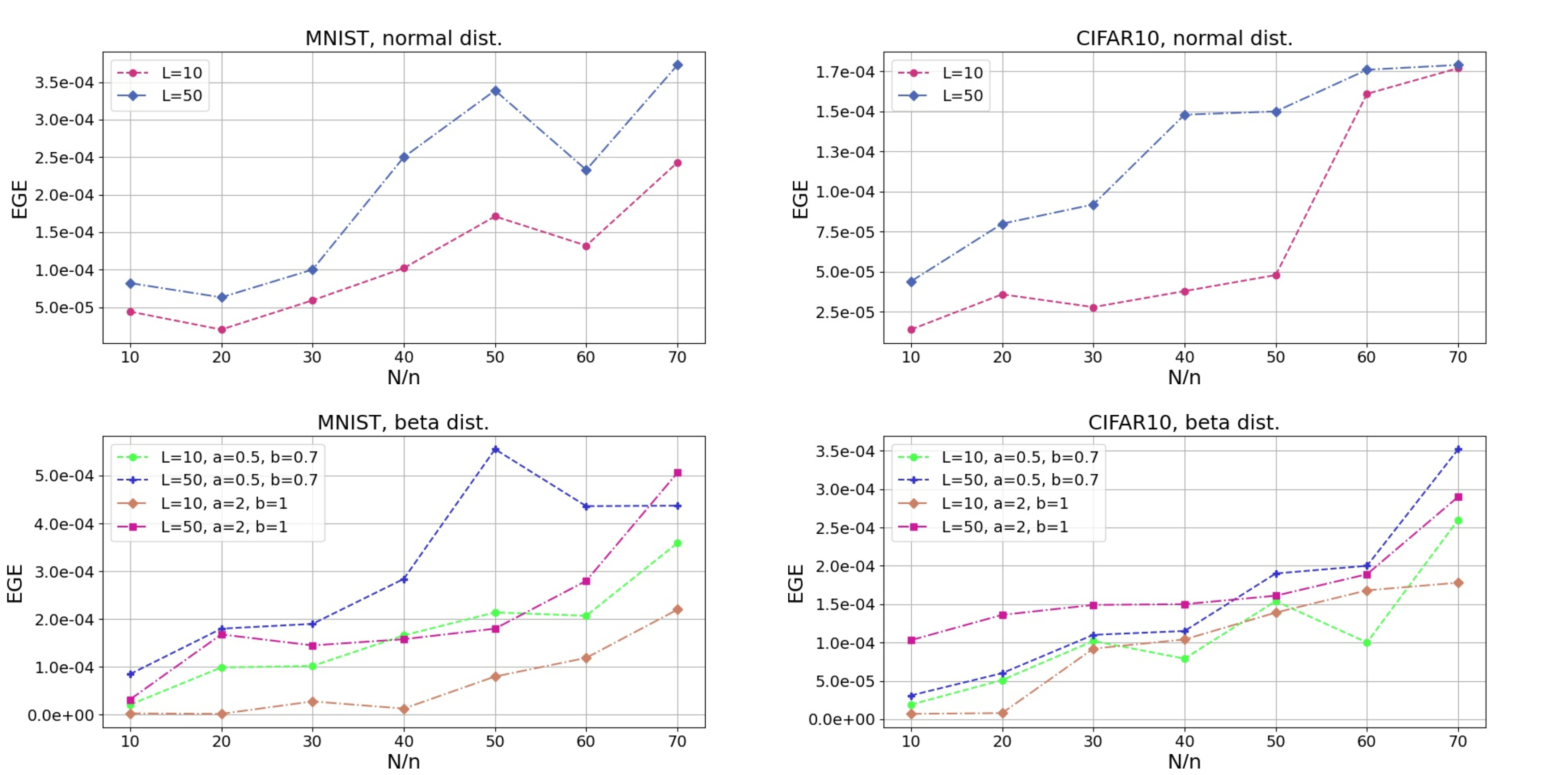}
    \captionsetup{justification=centering}
    \caption{Empirical generalization errors with normal (top) and Beta (bottom) distributions}
    \label{egecollage}
    \end{subfigure}
    \caption{Performance plots for 10- and 50-layer DECONET, $m=n/4$, tested on MNIST (left) and CIFAR10 (right) datasets.}
    \label{perfomplots}
\end{figure*}

\begin{figure}[h!]
\centering
\includegraphics[width=\linewidth]{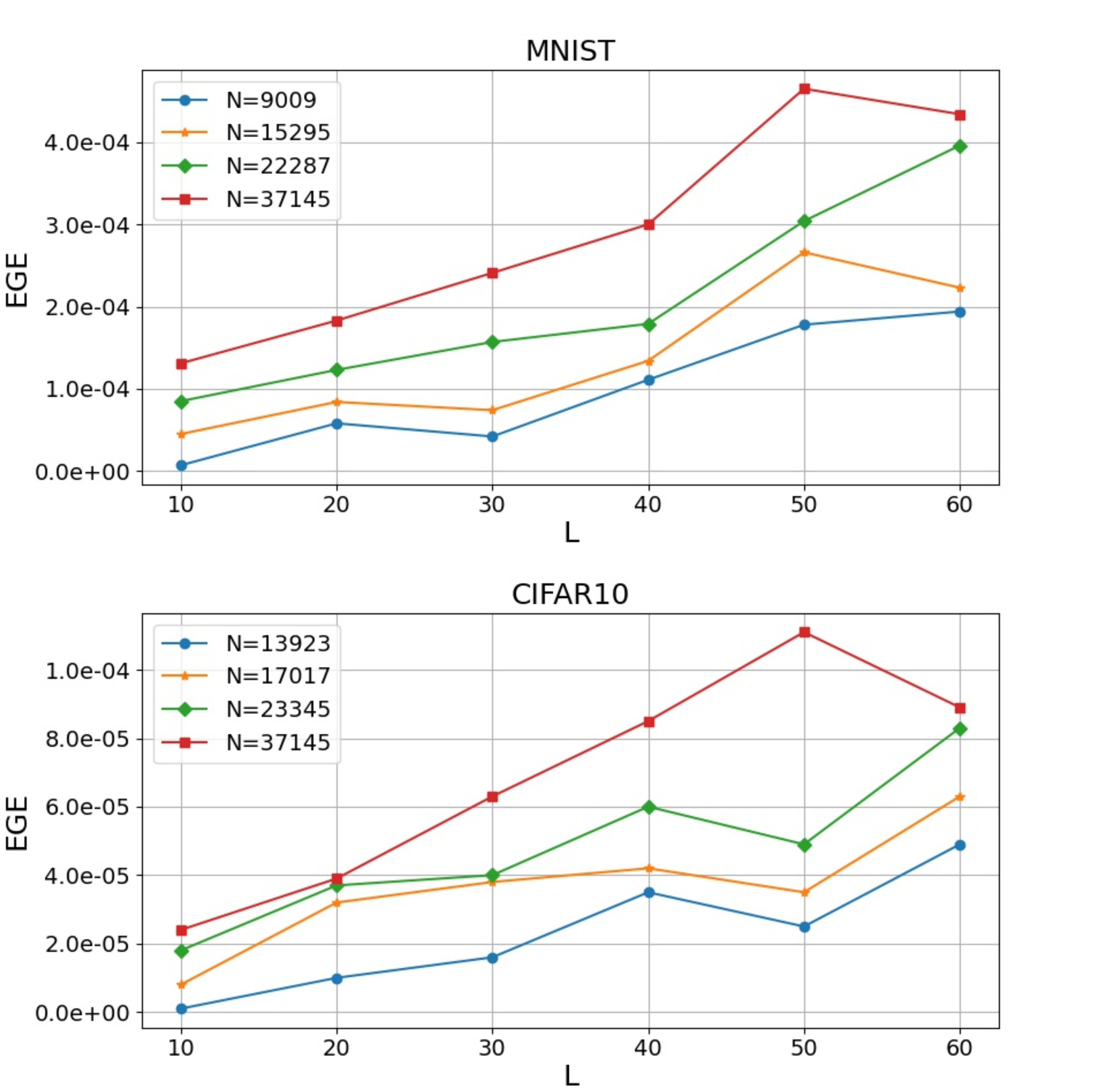}
\caption{Performance plots for DECONET with $50\%$ CS ratio, tested on MNIST (top) and CIFAR10 (bottom) datasets.}
\label{Lplots}
\end{figure}

\begin{figure}[t!]
\centering
\includegraphics[width=\linewidth]{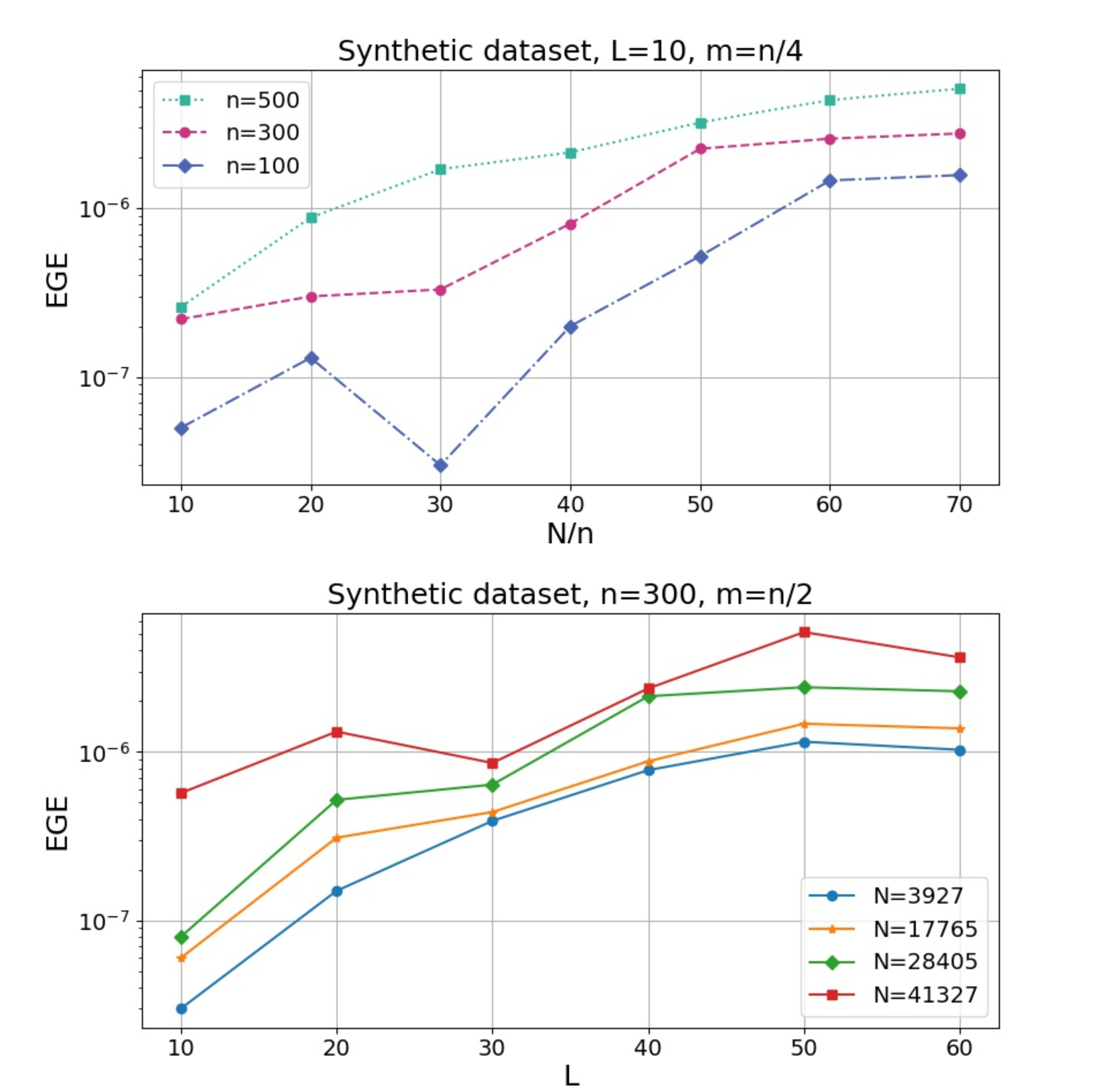}
\caption{Performance plots for DECONET, tested on a synthetic dataset, under different settings.}
\label{synthetic}
\end{figure}

\begin{table}[b]
    \centering
    \scalebox{0.9}{\begin{tabular}{||c|c|c|c||}
    \hline
     &  \multicolumn{3}{|c|}{Test MSE}\\
    \hline
    \diagbox{Method}{Dataset} & MNIST & CIFAR10 & Synthetic\\
    \hline\hline
    ACF (Wavelet) & 0.2515 & 0.3174 & 0.1585\\
    \hline
    ACF (TV) & 0.1978 & 0.3141 & 0.1051\\
    \hline
    DECONET (learnable) & \bf 0.0571 & \bf 0.0298 & $\mathbf{0.7523\cdot10^{-2}}$ \\
    \hline
    \end{tabular}}
    \caption{Comparison of MSEs achieved by DECONET with its learnable analysis operator, ACF with a redundant Haar wavelet transform and ACF with a total variation operator, on all datasets, with 10 layers/iterations. Bold letters indicate the best performance among three methods.}
    \label{acfcomp}
\end{table}

\section{Numerical Experiments}\label{expsec}
In this Section, we examine whether our theory regarding the generalization error of DECONET is consistent to real-world applications of our framework.
\subsection{Experimental Setup}
We train and test DECONET on a synthetic dataset of random vectors, drawn from the normal distribution (70000 training and 10000 test examples), and two real-world image datasets: MNIST (60000 training and 10000 test $28\times28$ image examples) and CIFAR10 (50000 training and 10000 test $32\times32$ coloured image examples). For the CIFAR10 dataset, we transform the images into grayscale ones. For both datasets, we consider the vectorized form of the images. We examine DECONET with varying number of layers $L$. We consider two CS ratios, i.e. $m/n=25\%$ and $m/n=50\%$. We choose a random Gaussian measurement matrix $A\in\mathbb{R}^{m\times n}$ and appropriately normalize it, i.e., $\Tilde{A}=A/\sqrt{m}$. We add zero-mean Gaussian noise $e$ with standard deviation $\mathrm{std}=10^{-4}$ to the measurements $y$, so that $y=\tilde{A}x+e$. We set $\varepsilon=\|y-\tilde{A}x\|_2$ and $x_0=A^Ty$, which are standard algorithmic setups. We take different values of $N$ and perform two different initializations for $W\in\RNn$: normal initialization and initialization based on Beta distribution \cite{dyingrelu}, with varying values of Beta's parameters $a$ and $b$. We set $\mu=100$, initial step sizes $t_0^1=t_0^2=1$ and step size multiplier $\theta_0=1$. For $t_k^1$, $t_k^2$, $\theta_k$, we apply the following update rules: $t_{k}^1=\alpha t_{k-1}^1$, $t_k^2=\beta t_{k-1}^2$, $\theta_k=\theta_{k-1}\cdot\theta'$, $k=1,\dots,L$, where $(\alpha,\beta)\in(0,1)\times(0,1)$, $\theta'=\frac{1-\sqrt{\mu/\Tilde{L}}}{1+\sqrt{\mu/\Tilde{L}}}$, respectively, and $\Tilde{L}$ is an upper bound on the smoothing parameter $\mu$; we set $\Tilde{L}=1000$. All networks are implemented in PyTorch \cite{pytorch} and trained using \emph{Adam} \cite{adam} algorithm, with batch size $128$. Adam constitutes a stochastic optimization method which can adaptively estimate lower-order moments of the gradient of \eqref{trainmse}. We employ the Pytorch implementation of Adam and set the initial learning rates to $\eta_\text{M}=10^{-2}$, $\eta_\text{C}=10^{-3}$ and $\eta_\text{S}=10^{-4}$, for the MNIST, CIFAR10 and synthetic datasets, respectively; the rest of Adam's parameters are set to their default values. For our experiments, we report the \emph{test MSE} defined by
\begin{equation}\label{testmse}
    \mathcal{L}_{test}=\frac{1}{d}\sum_{i=1}^d\|h(\tilde{y}_i)-\tilde{x}_i\|_2^2,
\end{equation}
where $\mathbf{D}=\{(\tilde{y}_i,\tilde{x}_i)\}_{i=1}^d$ 
is a set of $d$ test data, not used in the training phase. We also report the \emph{empirical generalization error} (EGE) defined by
\begin{equation}\label{genmse}
    \mathcal{L}_{gen}=|\mathcal{L}_{test}-\mathcal{L}_{train}|,
\end{equation}
where $\mathcal{L}_{train}$ is the train MSE defined in \eqref{trainmse}. Since test MSE approximates the true loss, we use \eqref{genmse} -- which can be explicitly computed -- to approximate the generalization error of \eqref{generror}. We train all networks, on all datasets, employing an early stopping technique \cite{earlystop} with respect to \eqref{genmse}. We repeat all the experiments at least 10 times and average the results over the runs.
We compare the reconstruction quality offered by DECONET, to the MSE achieved by the original optimization algorithm\footnote{One of the motivations for unrolling iterative schemes to corresponding neural networks relies on the fact that the latter yield a smaller reconstruction error, than their original iterative schemes \cite{ampnet}}, namely ACF (see Section \ref{acfsec}), with a similar structure. To that end, we choose as predefined sparsifiers a redundant Haar wavelet transform and a finite difference operator. The latter is associated to the popular method of total variation \cite{tv}, while redundant wavelets constitute standard choices of sparsifying transforms for analysis-based inverse problems \cite{genzel}, \cite{figu}. Finally, we compare the EGE of DECONET to the EGE achieved by two SotA unfolding networks serving as baselines: a recent variant of ISTA-net \cite{istagen} and ADMM-DAD net \cite{admmdad}. Both baselines jointly learn a decoder for CS and a sparsifying transform. Nevertheless, ISTA-net solves the CS problem employing synthesis sparsity, since the learnable sparsifier is orthogonal, while ADMM-DAD involves analysis sparsity, by learning a redundant analysis operator. Our choice of the aforementioned baselines is attributed to our interest of considering one SotA unfolding network from each sparsity ``category'', so that we examine a) how the EGE is affected by each of the two sparsity models and b) if DECONET outperforms\footnote{To the best of authors' knowledge, ADMM-DAD is the only unfolding network, apart from DECONET, that entails analysis sparsity for solving the CS problem} ADMM-DAD. For both baselines, we set the best hyper-parameters proposed by the original authors. From a complexity perspective, the most costly computation of both DECONET and ISTA-net consists in the multiplication of a matrix with its transpose ($N\times n$ with $n\times N$ for DECONET, $n\times m$ with $m\times n$ for ISTA-net), so that the dominant part per layer $L$ is of the order of $O(N^2n)$ and $O(n^2m)$, respectively. ADMM-DAD has a cubic complexity -- with respect to $n$ -- per layer $L$, since its ``heavier'' computation involves the inversion of a $n\times n$ matrix. Overall, the computational complexities of DECONET, ISTA-net and ADMM-DAD are of the order of $O(LN^2n)$, $O(Ln^2m)$ and $O(Ln^3)$, respectively.

\subsection{Experimental Results}
We test DECONET on all datasets under multiple experimental scenarios.

\subsubsection{Fixed CS ratio with varying $N/n$ and $L$, for different initializations, on image datasets}\label{exper1}
We examine the performance of 10- and 50-layer DECONET for a fixed $25\%$ CS ratio, varying redundancy ratio $N/n$ and both normal and Beta initializations for $W$. We report the results for MNIST and CIFAR10 in Fig. \ref{perfomplots}. As illustrated in Fig. \ref{testcollage}, the test MSEs, achieved by 10- and 50-layer DECONET on both datasets, drop as $L$ and $N/n$ increase, for both types of initialization. The decays seem reasonable, if one considers a standard analysis CS scenario: the reconstruction quality and performance of the analysis-$l_1$ algorithm, typically benefit from the (high) redundancy offered by the involved analysis operator, especially as the number of iterations/layers increases. Furthermore, Fig. \ref{egecollage} demonstrates that the EGE of DECONET increases as both $L$ and $N/n$ increase, for both normal and Beta initialization. For the latter, we observe that the different values of its parameters affect the generalization ability of DECONET on both image datasets. The overall performance of DECONET confirms our theoretical results depicted in Section \ref{gensec}, since EGEs seem to scale like $\sqrt{NL}$.
\begin{table*}[h!]
    \centering
    \scalebox{0.95}{\begin{tabular}{||c|c|c|c|c|c|c|c|c||}\hline
         &  \multicolumn{8}{|c|}{$25\%$ CS ratio}\\
         \hline
         Dataset & \multicolumn{3}{|c|}{MNIST} & \multicolumn{3}{|c|}{CIFAR10} \\
         \hline
         \diagbox{Decoder}{Layers} & $L=10$ & $L=20$ & $L=30$ & $L=10$ & $L=20$ & $L=30$ \\ \hline
         DECONET & \bf 0.000033 & \bf 0.000110 & \bf 0.000314 & \bf 0.000066 & \bf 0.000058 & \bf 0.000100\\
         \hline
         ADMM-DAD & 0.000212 & 0.000284 & 0.000394 & 0.000086 & 0.000102 & 0.000168 \\
         \hline
         ISTA-net & 0.013408 & 0.015099 & 0.007874 & 0.007165 & 0.005045 & 0.004120\\
         \hline
    \end{tabular}}
    
    \medskip
    
    \scalebox{0.95}{\begin{tabular}{||c|c|c|c|c|c|c|c|c||}\hline
         &  \multicolumn{7}{|c|}{$50\%$ CS ratio}\\
         \hline
         Dataset & \multicolumn{3}{|c|}{MNIST} & \multicolumn{3}{|c|}{CIFAR10} \\
         \hline
         \diagbox{Decoder}{Layers} & $L=10$ & $L=20$ & $L=30$ & $L=10$ & $L=20$ & $L=30$ \\ \hline
         DECONET & \bf 0.000131 & \bf 0.000183 & \bf 0.000241 & \bf 0.000024 & \bf 0.000039 & \bf 0.000063\\
         \hline
         ADMM-DAD & 0.000224 & 0.000455 & 0.000332 & 0.000094 & 0.000046 & 0.000087\\
         \hline
         ISTA-net & 0.016547 & 0.011624 & 0.009311 & 0.009274 & 0.006739 & 0.005576\\
         \hline
    \end{tabular}}
    \caption{Empirical generalization errors for 10-, 20- and 30-layer decoders, with $25\%$ and $50\%$ CS ratios, and fixed $N=37145$ for DECONET's and ADMM-DAD's sparsifiers. Bold letters indicate the best performance among the three decoders.}
    \label{decoders}
\end{table*}

\begin{figure}[h!]
    \centering
    \includegraphics[width=\linewidth]{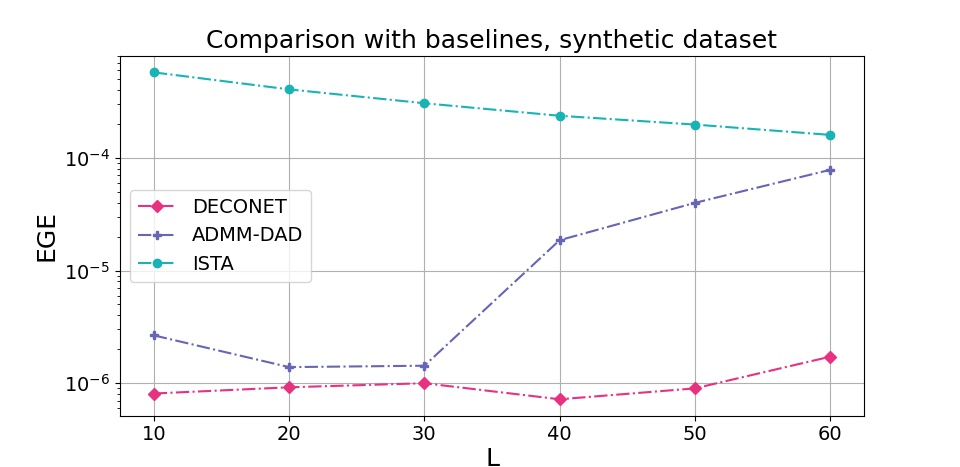}
    \caption{Performance plot for all decoders, with fixed $m,\,n,\,N$, on a synthetic dataset.}
    \label{baselines}
\end{figure}

\subsubsection{Fixed CS ratio, with varying $N$ and $L$, on image datasets}\label{exper2}
We examine the generalization ability of DECONET for $m=n/2$, with increasing number of layers $L$, under different choices of $N$ and normal initialization. Inspired by frames with redundancy ratio $N/n\notin\mathbb{N}$ \cite{shearletsframes}, we consider $N$ of the form
\begin{equation}\label{Nformula}
    N=pn+q,\qquad p,q\in\mathbb{N}.
\end{equation}
We report the results in Fig. \ref{Lplots} for MNIST and CIFAR10. Similarly to Section \ref{exper1}, we observe that the empirical generalization error increases in $L$ and $N$, for both datasets. Even though the upper bound in \eqref{genbound} depends on other terms too, the empirical generalization error appears to grow at the rate of $\sqrt{NL}$. The behaviour of DECONET again conforms with our theoretical results presented in Section \ref{mainsec}. One may also notice that -- in general -- we choose different $N$ for each of the two datasets. This is simply due to \eqref{Nformula}, i.e., $N$ depends on the vectorized ambient dimension $n$, which is different for each of the two datasets.

\subsubsection{Fixed CS ratio with varying $n$, $N$, $L$, on synthetic dataset}
Similarly to Sections \ref{exper1} and \ref{exper2}, we examine the generalization error that DECONET achieves on a synthetic dataset of random vectors, with a normally initialized $W$. We present the results in Fig.~\ref{synthetic} (and note that all plots regarding the synthetic data are made in logarithmic scale). Particularly, the top plot of Fig.~\ref{synthetic} demonstrates how the EGE for 10-layer DECONET scales, for different values of the ambient dimension $n$, with $25\%$ CS ratio, as $N/n$ increases. On the other hand, as depicted in the bottom plot of Fig. \ref{synthetic}, we revisit the experimental reasoning of Section \ref{exper2}, with fixed $n=300$ and $m=n/2$. In both subplots, we see that the EGEs achieved by DECONET comply with our theory.

\subsubsection{Comparison between DECONET and ACF}
We investigate the reconstruction quality -- on all datasets -- offered by DECONET with a learnable analysis operator satisfying $N/n=50$, to the test MSE achieved by ACF, with two different redundant analysis operators: wavelets and total variation operator. We set $m=n/4$, $L=10$ (for all methods) and fix $n=100$ for the synthetic dataset. We present our results in Table~\ref{acfcomp}, which showcases that the test MSE achieved by both instances of ACF is larger than DECONET's, consistently for all three datasets. This behaviour complies with the motivation for interpreting iterative methods as neural networks, since the latter are able to achieve a smaller reconstruction error than their iterative schemes, for the same number of layers/iterations.

\subsubsection{Comparison to baselines}
We examine how analysis and synthesis sparsity models affect the generalization ability of CS-oriented unfolding networks. Towards this end, we compare the proposed DECONET's decoder to ISTA-net's and ADMM-DAD's decoders. For the image datasets, the comparisons are made for 10, 20 and 30 layers, with $25\%$ and $50\%$ CS ratio, and fixed $N=37145$ for both DECONET's and ADMM-DAD's sparsifiers. For the synthetic dataset, we fix $N=12000$ (for both DECONET's and ADMM-DAD's sparsifiers), $n=300$, $m=n/2$, and vary $L$. We report the empirical generalization errors for the image datasets in Table~\ref{decoders} and for the synthetic dataset in Fig.~\ref{baselines}. We conjecture that DECONET should outperform ISTA-net in terms of EGE. Our speculation is attributed to similar (albeit limited) experimental findings in \cite[Section 4]{admmdad}, which indicate that analysis sparsity can affect the generalization ability of unfolding networks. This is indeed the case according to Table~\ref{decoders} and Fig.~\ref{baselines}, which demonstrate that our proposed decoder outperforms both baseline decoders, consistently for all datasets. In fact, both DECONET and ADMM-DAD  outperform the synthesis-sparsity-based ISTA-net. This behaviour showcases that learning a redundant sparsifier instead of an orthogonal one, improves the performance of a CS-oriented unfolding network. Additionally, our proposed network outperforms ADMM-DAD, which is considered to be a SotA unfolding network for analysis-based CS. Finally, our theoretical results on the generalization error of DECONET seem to align with the experiments, since the EGE of DECONET increases as $L$ also increases. Interestingly, we observe that the EGE of DECONET on the CIFAR10 dataset decreases as the number of measurements $m$ increases. This behaviour is not reflected by our theoretical results (which are independent of $m$), but it could serve as a potential line of future work.

\section{Conclusion and Future Work}
\label{concsec}
In this paper we derived DECONET, a new deep unfolding network for solving the analysis-sparsity-based Compressed Sensing problem. DECONET jointly learns a decoder for CS and a redundant sparsifying analysis operator. Furthermore, we estimated the generalization error of DECONET, in terms of the Rademacher complexity of the associated hypothesis class. Our generalization error bounds roughly scale like the square root of the product between the number of layers and the redundancy of the learnable sparsifier. To the best of our knowledge, this is the first result of its kind for unfolding networks solving the analysis-based CS problem. Furthermore, we conducted experiments that confirmed the validity of our theoretical results and compared DECONET to state-of-the-art CS-oriented unfolding networks. Our proposed network outperformed the baselines, consistently for synthetic and real-world datasets. As a future direction, we would like to examine the performance of our proposed framework on speech datasets and experiment with different values of the non-learnable parameters. Additionally, it would be interesting to further characterize (e.g. in terms of structure) the sparsifying transform that DECONET learns. Last but not least, it would be intriguing to investigate the tightness of our delivered generalization error bounds. For example, we could employ techniques presented in \cite{ecover}, \cite{mutualinf}, in order to examine whether our bounds could be independent of the number of layers and/or the redundancy of the learnable sparsifying transform.

\section*{Acknowledgements}
\noindent V. Kouni would like to thank Holger Rauhut for his advice and helpful conversations around the subject covered in this paper.
\printbibliography

\vskip -3\baselineskip plus -5fil
\begin{IEEEbiographynophoto}{Vicky Kouni}
received her B.Sc. and M.Sc. in Mathematics, from the Dep. of Mathematics, National \& Kapodistrian University of Athens, Greece. She is currently a PhD student at the Dep. of Informatics \& Telecommunications, National \& Kapodistrian University of Athens, Greece. She has received scholarships and awards for her studies and research, including the recent \emph{Scholarship for Senior Doctorate Students} by the greek State Scholarships Foundation (IKY). Her research interests are mainly focused on deep unfolding, learning theory, compressed sensing, sparse representations, applied harmonic analysis, and their applications in audio and image processing.
\end{IEEEbiographynophoto}
\vskip -1\baselineskip plus -1fil
\begin{IEEEbiographynophoto}{Yannis Panagakis}
received his PhD and MSc
from the Dep. of Informatics, Aristotle
University of Thessaloniki, and his B.Sc.
in Informatics \& Telecommunications from the
National \& Kapodistrian University of Athens,
Greece. He previously held research and academic positions at the Samsung AI Centre, Cambridge, U.K., Middlesex University London, and Imperial College London, U.K. He is currently an Associate Professor of machine learning and signal processing in the Dep. of Informatics \& Telecommunications, National \& Kapodistrian University of Athens,
Greece. He has received prestigious scholarships and awards for his studies and research, including the Marie-Curie Fellowship in 2013. His research interests include: machine learning and its interface with tensor methods, deep learning, computer vision, signal processing and mathematical optimization. He has published over 80 articles in leading journals and conferences.
\end{IEEEbiographynophoto}

\end{document}


\maketitle

\section*{Code availability}
For reproducibility purposes, our code regarding the experiments presented in the accompanying paper is publicly available at \url{https://github.com/vicky-k-19/deconet}.

\section*{Proof of Lemma \ref{g12bound}}

\begin{proof}
By definition of the matrices $G_k^1$, $G_k^2$, we get
\begin{align*}
    \|G^1_{k}\|_{\opnorm}&+\|G^2_{k}\|_{\opnorm}\\
    \leq&\bigg[\theta_{k}\|I-\theta^{-1}_{k}\mu^{-1}t^1_{k}WW^T\|_{2\rightarrow2}\\
   &+\mu^{-1}t^1_{k}\|W\|_{2\rightarrow2}\|A\|_{\opnorm}\\
   &+(1-\theta_{k})\|I-\theta^{-1}_{k}\mu^{-1}t^1_{k}WW^T\|_{2\rightarrow2}\notag\\
   &+(1-\theta_{k})\theta^{-1}_{k}\mu^{-1}t^1_{k}\|W\|_{2\rightarrow2}\|A\|_{\opnorm}\bigg]\\
   &+\bigg(\mu^{-1}t^2_{k}\|A\|_{2\rightarrow2}\|W\|_{2\rightarrow2}\\
   &+\theta_{k}\|I-\theta^{-1}_{k}\mu^{-1}t^2_{k}AA^T\|_{2\rightarrow2}\\
   &+(1-\theta_{k})\theta^{-1}_{k}\mu^{-1}t^2_{k}\|A\|_{2\rightarrow2}\|W\|_{2\rightarrow2}\\
   &+(1-\theta_{k})\|I-\theta^{-1}_{k}\mu^{-1}t^2_{k}AA^T\|_{2\rightarrow2}\bigg)\notag\\
   =&\|I-\theta^{-1}_{k}\mu^{-1}t^1_{k}WW^T\|_{2\rightarrow2}\\
   &+\|I-\theta^{-1}_{k}\mu^{-1}t^2_{k}AA^T\|_{2\rightarrow2}+\|W\|_{2\rightarrow2}\|A\|_{2\rightarrow2}\\
   &\cdot\big(\theta^{-1}_{k}\mu^{-1}t_{k}^1+\mu^{-1}t_{k}^1+\theta^{-1}_{k}\mu^{-1}t^2_{k}+\mu^{-1}t^2_{k}\big).
\end{align*}
We set $c_{1,k}=\theta^{-1}_{k}\mu^{-1}t^1_{k}$, $c_{2,k}=\theta^{-1}_{k}\mu^{-1}t^2_{k}$ and by definition of $\mu$, $t^1_{k},\, t^2_{k},\,\theta_k$, we obtain $\{c_{1,k}\}_{k\geq0}\leq1$, $\{c_{2,k}\}_{k\geq0}\leq1$. 
We also have
\begin{align*}
    \|I-c_{1,k}WW^T\|_{2\rightarrow2}&\leq\max\{{c_{1,k}\Lambda^2}-1,1\}\leq{c_{1,k}\Lambda^2}\\
    \|I-c_{2,k}AA^T\|_{2\rightarrow2}&\leq\max\{{c_{2,k}\|A\|_{\opnorm}^2}-1,1\}\\
    &\leq{c_{2,k}\|A\|_{\opnorm}^2},
\end{align*}
so that
\begin{align*}
    2\|G^1_{k}\|_{\opnorm}+2\|G^2_{k}\|_{\opnorm}+1\leq&2\big[c_{1,k}\Lambda^2+c_{2,k}\|A\|_{\opnorm}^2\\
    &+\|A\|_{\opnorm}\Lambda(c_{1,k}+\theta_k c_{1,k}\\
    &+c_{2,k}+\theta_k c_{2,k})\big]+1.
\end{align*}
Since $\theta_k\leq1$, we obtain 
\begin{align*}
    2&\|G^1_{k}\|_{\opnorm}+2\|G^2_{k}\|_{\opnorm}+1\\ &\leq2\big[c_{1,k}\Lambda^2+c_{2,k}\|A\|_{\opnorm}^2+2\|A\|_{\opnorm}\Lambda(c_{1,k}+c_{2,k})\big]+1\\&:=\Gamma_k.
\end{align*}
The assumptions $c_{1,k}\Lambda\leq1$, $c_{1,k}\Lambda^2\leq1,\,c_{2,k}\|A\|_{\opnorm}^2\leq1$ yield
\begin{align*}
    \Gamma_k\leq&2\big[2+2\|A\|_{\opnorm}\Lambda c_{1,k}+2\|A\|_{\opnorm}\Lambda c_{2,k}]+1\notag\\
    &\overset{\|A\|_{\opnorm}\approx1}{\leq}4\big[1+\|A\|_{\opnorm}+\Lambda]+1:=\gamma
\end{align*}
and the proof follows.
\end{proof}

\section*{Proof of Theorem \ref{perturbation}}

\begin{proof}
First, we set $f^0_{W_1}(Y)=f^0_{W_2}(Y)=Y$ for a uniform treatment of all layers. Then, we write $\{G_{1,k}^i\}_{i=1,2},\,\{G_{2,k}^i\}_{i=1,2}$ (similarly for $\{B^i_{1,k}\}_{i=1,2}$, $\{B^i_{2,k}\}_{i=1,2}$) to denote the dependency on $W_1,W_2$, respectively. By definition of the soft-thresholding and truncation operators, $\mathcal{S}(\cdot,\cdot)$ and $\mathcal{T}(\cdot,\cdot)$ are 1-Lipschitz functions with respect to the first parameter. Due to the Lipschitzness of $\mathcal{S}(\cdot,\cdot)$ and $\mathcal{T}(\cdot,\cdot)$, the estimates $\|D_k\|_{2\rightarrow2}\leq1$ and $\|\Theta_k\|_{2\rightarrow2}=1$ that hold for any $k\geq0$, and the introduction of mixed terms, we get
\begin{align*}
    \|f^k_{W_1}(Y)&-f^k_{W_2}(Y)\|_F\leq\|D_{k-1}f^{k-1}_{W_1}(Y)+\Theta_{k-1}\sigma(f_{W_1}^{k-1})\\
    &-D_{k-1}f^{k-1}_{W_2}(Y)-\Theta_{k-1}\sigma(f_{W_2}^{k-1})\|_F\\
    \leq&\|D_{k-1}\|_{2\rightarrow2}\|f^{k-1}_{W_1}(Y)-f^{k-1}_{W_2}(Y)\|_F\\
    &+\|\Theta_{k-1}\|_{2\rightarrow2}\|\sigma(f_{W_1}^{k-1})-\sigma(f_{W_2}^{k-1})\|_F\\
    \leq&\|f^{k-1}_{W_1}(Y)-f^{k-1}_{W_2}(Y)\|_F+2\|G_{1,k-1}^1f_{W_1}^{k-1}(Y)\\
    &-B_{1,k-1}^1-G_{2,k-1}^1f_{W_2}^{k-1}(Y)\\
    &+B_{2,k-1}^1\|_F+2\|G_{1,k-1}^2f_{W_1}^{k-1}(Y)\\
    &-B_{1,k-1}^2-G_{2,k-1}^2f_{W_2}^{k-1}(Y)+B_{2,k-1}^2\|_F\\
    =&\|f^{k-1}_{W_1}(Y)-f^{k-1}_{W_2}(Y)\|_F\\
    &+2\|G_{1,k-1}^1f_{W_1}^{k-1}(Y)-B_{1,k-1}^1+G_{1,k-1}^1f_{W_2}^{k-1}(Y)\\
    &-G_{1,k-1}^1f_{W_2}^{k-1}(Y)-G_{2,k-1}^1f_{W_2}^{k-1}(Y)+B_{2,k-1}^1\|_F\\
    &+2\|G_{1,k-1}^2f_{W_1}^{k-1}(Y)-B_{1,k-1}^2+G_{1,k-1}^2f_{W_2}^{k-1}(Y)\\
    &-G_{1,k-1}^2f_{W_2}^{k-1}(Y)-G_{2,k-1}^2f_{W_2}^{k-1}(Y)+B_{2,k-1}^2\|_F\\
    \leq&\|f^{k-1}_{W_1}(Y)-f^{k-1}_{W_2}(Y)\|_F\\
    &+2(\|G_{1,k-1}^1\|_{\opnorm}\|f^{k-1}_{W_1}(Y)-f^{k-1}_{W_2}(Y)\|_F\\
    &+\|f_{W_2}^{k-1}(Y)\|_F\|G_{2,k-1}^1-G_{1,k-1}^1\|_{\opnorm}\\
    &+\|B_{2,k-1}^1-B_{1,k-1}^1\|_F)\\
    &+2(\|G_{1,k-1}^2\|_{\opnorm}\|f^{k-1}_{W_1}(Y)-f^{k-1}_{W_2}(Y)\|_F\\
    &+\|f_{W_2}^{k-1}(Y)\|_F\|G_{2,k-1}^2-G_{1,k-1}^2\|_{\opnorm}\\
    &+\underbrace{\|B_{2,k-1}^2-B_{1,k-1}^2\|_F}_{=0\text{, since it is not parameter-dependent}}).
\end{align*}
Consequently,
\begin{align*}
    \|f^k_{W_1}(Y)-&f^k_{W_2}(Y)\|_F\\
    \leq&\|f^{k-1}_{W_1}(Y)-f^{k-1}_{W_2}(Y)\|_F(2\|G_{1,k-1}^1\|_{\opnorm}\\
    &+2\|G_{1,k-1}^2\|_{\opnorm}+1)+2\|f^{k-1}_{W_2}(Y)\|_F\\
    &\cdot(\|G_{1,k-1}^2-G_{2,k-1}^2\|_{\opnorm}\\
    &+\|G_{1,k-1}^1-G_{2,k-1}^1\|_{\opnorm})\\
    &+2\|B_{2,k-1}^1-B_{1,k-1}^1\|_F.
\end{align*}
For simplification, we treat separately the terms appearing in the last estimate.\\
\underline{1. Estimation of $2\|G_{1,k-1}^1\|_{\opnorm}+2\|G_{1,k-1}^2\|_{\opnorm}+1$:}\\

\noindent Due to Lemma \ref{g12bound} and the assumption that $W_1,W_2\in\mathcal{B}_\Lambda$, we have
\begin{equation*}
    2\|G_{1,k-1}^1\|_{\opnorm}
    +2\|G_{1,k-1}^2\|_{\opnorm}+1\leq\Gamma_{k-1},
\end{equation*}
with $\{\Gamma_k\}_{k\geq0}$ defined as in Lemma \ref{g12bound}.\\

\noindent\underline{2. Estimation of}\\
\underline{ $\|G^1_{2,k-1}-G^1_{1,k-1}\|_{\opnorm}+\|G^2_{2,k-1}-G^2_{1,k-1}\|_{\opnorm}$:}
\begin{align*}
    \|G^1_{2,k-1}&-G^1_{1,k-1}\|_{\opnorm}+\|G^2_{2,k-1}-G^2_{1,k-1}\|_{\opnorm}\\
    \leq&\big[\mu^{-1}t^1_{k-1}\|W_2W_2^T-W_1W^T_1\|_{2\rightarrow2}\\
    &+\mu^{-1}t^1_{k-1}\|A\|_{2\rightarrow2}\|W_2-W_1\|_{2\rightarrow2}\\
    &+(1-\theta_{k-1})\theta^{-1}_{k-1}\mu^{-1}t^1_{k-1}\|W_2W_2^T-W_1W^T_1\|_{2\rightarrow2}\\
    &+(1-\theta_{k-1})\theta^{-1}_{k-1}\mu^{-1}t^1_{k-1}\|A\|_{2\rightarrow2}\|W_2-W_1\|_{2\rightarrow2}\big]\\
    &+\big[\mu^{-1}t^2_{k-1}\|A\|_{2\rightarrow2}\|W_2-W_1\|_{2\rightarrow2}\\
    &+(1-\theta_{k-1})\theta^{-1}_{k-1}\mu^{-1}t^2_{k-1}\|A\|_{2\rightarrow2}\|W_2-W_1\|_{2\rightarrow2}\big]
\end{align*}
Similarly to Lemma \ref{g12bound}, we set $c_{1,k}=\theta^{-1}_{k}\mu^{-1}t^1_{k}$, $c_{2,k}=\theta^{-1}_{k}\mu^{-1}t^2_{k}$, for all $k\geq0$, with $c_{1,-1}=c_{2,-1}=0$. The combination of the previous statements and the assumption that $0<\{\theta_k\}_{k\geq0}\leq1$, with $\theta_0=\theta_{-1}=1$ , yield
\begin{multline*}
    \|G^1_{2,k-1}-G^1_{1,k-1}\|_{\opnorm}+\|G^2_{2,k-1}-G^2_{1,k-1}\|_{\opnorm}\\
    \leq c_{1,k-1}\|W_2W_2^T-W_1W^T_1\|_{2\rightarrow2}\\
    +\|A\|_{\opnorm}\|W_2-W_1\|_{2\rightarrow2}(c_{1,k-1}+c_{2,k-1}).
\end{multline*}
Furthermore, since $\|W_1\|_{2\rightarrow2},\,\|W_2\|_{2\rightarrow2}\leq\Lambda$, we obtain
\begin{multline*}
    \|W_2W_2^T-W_1W^T_1\|_{2\rightarrow2}\\
    =\|W_2W_2^T-W_1W_2^T+W_1W_2^T-W_1W^T_1\|_{2\rightarrow2}\\
    \leq\|W_2\|_{2\rightarrow2}\|W_2-W_1\|_{2\rightarrow2}+\|W_1\|_{2\rightarrow2}\|W_2-W_1\|_{2\rightarrow2}\implies\\
    \|W_2W_2^T-W_1W^T_1\|_{2\rightarrow2}\leq2\Lambda\|W_2-W_1\|_{2\rightarrow2}.
\end{multline*}

\noindent With the previous estimate at hand, we get
\begin{align*}
    \|G^1_{2,k-1}-&G^1_{1,k-1}\|_{\opnorm}+\|G^2_{2,k-1}-G^2_{1,k-1}\|_{\opnorm}\\
    \leq&2\Lambda c_{1,k-1}\|W_2-W_1\|_{2\rightarrow2}\\
    &+\|A\|_{2\rightarrow2}\|W_2-W_1\|_{2\rightarrow2}(c_{1,k-1}+c_{2,k-1})\implies\\
    \|G^1_{2,k-1}-&G^1_{1,k-1}\|_{\opnorm}+\|G^2_{2,k-1}-G^2_{1,k-1}\|_{\opnorm}\\
    \leq&(2\Lambda c_{1,k-1}+\|A\|_{2\rightarrow2}\\
    &\cdot(c_{1,k-1}+c_{2,k-1}))\|W_2-W_1\|_{2\rightarrow2}.
\end{align*}

\noindent\underline{3. Estimation of $\|B_{2,k-1}^1-B_{1,k-1}^1\|_F$}.
\begin{multline*}
    \|B_{2,k-1}^1-B_{1,k-1}^1\|_F\leq\theta^{-1}_{k-1}t^1_{k-1}\|X_0\|_F\|W_2-W_1\|_{2\rightarrow2}\\
    \leq\mu c_{1,k-1}\|A\|_{\opnorm}\|Y\|_F\|W_2-W_1\|_{2\rightarrow2}.
\end{multline*}
Now, we combine estimates from 1., 2., 3. and apply Lemma \ref{fbound}, to get
\begin{align*}
    \|f^k_{W_2}(Y)-&f^k_{W_1}(Y)\|_F\leq\Gamma_{k-1}\|f^{k-1}_{W_2}(V)-f^{k-1}_{W_1}(V)\|_F\\
    &+\bigg(2\|f_{W_2}^{k-1}(V)\|_F\big[2\Lambda c_{1,k-1}+\|A\|_{2\rightarrow2}(c_{1,k-1}\\
    &+c_{2,k-1})\big]+2\mu c_{1,k-1}\|X_0\|_F\bigg)\|W_2-W_1\|_{2\rightarrow2}\\
    \leq&\Gamma_{k-1}\|f^{k-1}_{W_2}(V)-f^{k-1}_{W_1}(V)\|_F+\big[2\Delta_{k-1}\\
    &\cdot(2\Lambda c_{1,k-1}+\|A\|_{2\rightarrow2}(c_{1,k-1}+c_{2,k-1}))\\
    &+2\mu c_{1,k-1}\|A\|_{\opnorm}\|Y\|_F\big]\|W_2-W_1\|_{2\rightarrow2},
\end{align*}
where
\begin{multline*}
    \Delta_k=2\mu\|Y||_F\Bigg[\sum_{i=0}^{k-1}\Bigg(\bigg(\|A\|_{\opnorm}(c_{1,i-1}\Lambda\\
    +c_{2,i-1}\|A\|_{\opnorm})+c_{2,i-1}\bigg)\prod_{j=i}^{k-1}\Gamma_j\Bigg)\\
    +\|A\|_{\opnorm}(c_{1,k-1}\Lambda+c_{2,k-1}\|A\|_{\opnorm})+ c_{2,k-1}\Bigg],
\end{multline*}
with $\Delta_0=0$. Now, we set
\begin{align*}
    E_k=&2\Delta_{k-1}(2\Lambda c_{1,k-1}+\|A\|_{2\rightarrow2}(c_{1,k-1}+c_{2,k-1}))\\
    &+2\mu c_{1,k-1}\|A\|_{\opnorm}\|Y\|_F,
\end{align*}
thus
\begin{align*}
    \|f^k_{W_2}(Y)-f^k_{W_1}(Y)\|_F\leq&\Gamma_{k-1}\|f^{k-1}_{W_2}(V)-f^{k-1}_{W_1}(V)\|_F\notag\\
    &+E_k\|W_2-W_1\|_{2\rightarrow2}.
\end{align*}
Using the abbreviations defined by $\Gamma_k,\,\Delta_k,\,E_k$, the general formula for $K_L$ is
\begin{align*}
    K_L=\sum_{k=1}^L\left(\max_{0\leq i\leq L-1}\Gamma_i\right)^{L-k}E_k\qquad\text{for}\quad L\geq1.
\end{align*}
We can now prove that $f_W^L$ is Lipschitz continuous w.r.t. $W$, for any number of layers $L\geq1$, with Lipschitz constants $K_L$ given by the aforementioned general formula. In order to do so, we employ the fact that
\begin{align*}
    \|f^k_{W_2}(Y)-f^k_{W_1}(Y)\|_F\leq&\Gamma_{k-1}\|f^{k-1}_{W_2}(V)-f^{k-1}_{W_1}(V)\|_F\notag\\
    &+E_k\|W_2-W_1\|_{2\rightarrow2}.
\end{align*}
For $L=1$, we can directly calculate $K_1$:
\begin{multline*}
    \|f^1_{W_2}(Y)-f^1_{W_1}(Y)\|_F\leq2t_0^1\|A\|_{\opnorm}\|Y\|_F\|W_2-W_1\|_{\opnorm}\\
    =2\underbrace{\theta_0^{-1}t_0^1}_{=1}\|A\|_{\opnorm}\|Y\|_F\|W_2-W_1\|_{\opnorm}\\
    =2\mu c_{1,0}\|A\|_{\opnorm}\|Y\|_F\|W_2-W_1\|_{\opnorm},
\end{multline*}
so that $2\mu c_{1,0}\|A\|_{\opnorm}\|Y\|_F=E_1=K_1$, as claimed in the general formula of $K_L$.
Let us assume that $f_W^L$ is Lipschitz continuous w.r.t. $W$, for some $L\in\mathbb{N}$. Then, for $L+1$:
\begin{align*}
    \|f^{L+1}_{W_2}(Y)-&f^{L+1}_{W_1}(Y)\|_F\\
    \leq&\Gamma_L\|f^L_{W_2}(Y)-f^L_{W_1}(Y)\|_F+E_{L+1}\|W_2-W_1\|_F\\
    \leq&(\Gamma_LK_L+E_{L+1})\|W_2-W_1\|_F\\
    \leq&\left((\max_{0\leq i\leq L}\Gamma_i)K_L+E_{L+1}\right)\|W_2-W_1\|_F\\
    =&\left((\max_{0\leq i\leq L}\Gamma_i)\sum_{k=1}^L(\max_{0\leq i\leq L-1}\Gamma_i)^{L-k}E_{k}+E_{L+1}\right)\\
    &\cdot\|W_2-W_1\|_F\\
    \leq&\left(\sum_{k=1}^L(\max_{0\leq i\leq L}\Gamma_i)^{k}E_{k}+(\max_{0\leq i\leq L}\Gamma_i)^0E_{L+1}\right)\\
    &\cdot\|W_2-W_1\|_F\\
    =&\left(\sum_{k=1}^{L+1}(\max_{0\leq i\leq L}\Gamma_i)^{L+1-k}E_{k}\right)\|W_2-W_1\|_F\\
    =&K_{L+1}\|W_2-W_1\|_F.
\end{align*}
We successfully calculated the desired $K_L$. Under the additional assumptions $c_{1,k}\Lambda\leq1$,  $c_{1,k}\Lambda^2\leq1,\,c_{2,k}\|A\|_{\opnorm}^2\leq1$, for all $k\geq0$, we may apply the second part of Lemma \ref{fbound} on $K_L$, to obtain a simplified upper bound on the latter. Thus, for $\gamma$ and $\zeta_k$ defined in Lemmata \ref{g12bound} and \ref{fbound} respectively, we have
\begin{align}
    K_L\leq&2\mu\|Y\|_F\Bigg[\mu^{-1}\|A\|_{\opnorm}+\sum_{k=2}^L\Bigg(2\gamma^{L-k}(\|A\|_{\opnorm}+1)\notag\\
    &\cdot(\|A\|_{\opnorm}+3)(\zeta_{k-1}+1)+\|A\|_{\opnorm}\Bigg)\Bigg]\notag\\
    =&2\mu\|Y\|_F\Bigg[\|A\|_{\opnorm}(L-1+\mu^{-1})\notag\\
    &+Z\gamma^L\sum_{k=2}^L\bigg(\gamma^{-k}(\zeta_{k-1}+1)\bigg)\Bigg]\notag,
\end{align}
where $Z=2(\|A\|_{\opnorm}+1)(\|A\|_{\opnorm}+3)$. Hence, we calculate the following sum:
\begin{align*}
    \gamma^L\sum_{k=2}^L\bigg(\gamma^{-k}&(\zeta_{k-1}+1)\bigg)=\gamma^L\sum_{k=2}^L\gamma^{-k}\left(\frac{\gamma^{k-1}-1}{\gamma-1}+1\right)\\
    &=\frac{\gamma^L}{\gamma-1}\sum_{k=2}^L\left(\frac{\gamma^L}{\gamma}-\frac{2}{\gamma^k}+\frac{\gamma}{\gamma^k}\right)\\
    &=\frac{\gamma^L}{\gamma-1}\left(\frac{L-1}{\gamma}+(\gamma-2)\sum_{k=2}^L\frac{1}{\gamma^k}\right)\\
    &=\frac{\gamma^L}{\gamma-1}\left(\frac{L-1}{\gamma}+(\gamma-2)\frac{\gamma^{1-L}-1}{\gamma^{-1}-1}\right)\\
    &=\frac{1}{\gamma-1}\left(\frac{\gamma^L(L-1)}{\gamma}+\frac{\gamma(\gamma-2)(\gamma^L-\gamma)}{\gamma-1}\right)\\
    &=\frac{\gamma^L(L-1)(\gamma-1)+\gamma^2(\gamma-2)(\gamma^L-\gamma)}{\gamma(\gamma-1)^2}\\
    &=\gamma^L\left(\frac{L-1}{\gamma(\gamma-1)}+\frac{\gamma(\gamma-2)}{(\gamma-1)^2}\right)-\frac{\gamma^2(\gamma-2)}{(\gamma-1)^2}.
\end{align*}
The proof is complete.
\end{proof}